# Review of important reactions for the nitrogen chemistry in the interstellar medium

V. Wakelam[1,2], I. W. M. Smith[3], J.-C. Loison[4,5], D. Talbi[6], S. J. Klippenstein[7], A. Bergeat[4,5], W. D. Geppert[8], K. M. Hickson[4,5]

[1] Univ. Bordeaux, LAB, UMR 5804, F-33270, Floirac, France
[2] CNRS, LAB, UMR 5804, F-33270, Floirac, France
[3] University Chemical Laboratories, Lensfield Road, Cambridge CB2 1EW, UK
[4] Univ. Bordeaux, ISM, CNRS UMR 5255, F-33400 Talence, France
[5] CNRS, ISM, CNRS UMR 5255, F-33400 Talence, France
[6] Université Montpellier II - GRAAL, CNRS - UMR 5024, place Eugène Bataillon, 34095 Montpellier, France
[7] Chemical Sciences and Engineering Division, Argonne National Laboratory, Argonne, IL, 60439, USA
[8] Department of Physics, University of Stockholm, Roslagstullbacken 21, S-10691 Stockholm

**Abstract**

Predictions of astrochemical models depend strongly on the reaction rate coefficients used in the simulations. We reviewed a number of key reactions for the chemistry of nitrogen-bearing species in the dense interstellar medium and proposed new reaction rate coefficients for those reactions. The details of the reviews are given in the form of a datasheet associated with each reaction. The new recommended rate coefficients are given with an uncertainty and a temperature range of validity and will be included in KIDA (http://kida.obs.u-bordeaux1.fr).

*Subject headings: Physical data and processes: astrochemistry — Astronomical databases: miscellaneous — ISM: abundances, molecules*

## 1. Introduction

The interpretation of the observation of molecular lines in the interstellar medium (as well as planetary atmospheres) require an assumption on the abundance and distribution of the molecule in the observed source. To obtain this information, astrophysicists often use chemical models to make predictions. Those models compute the chemical composition of the gas and dust as a function of time based on a number of physico-chemical processes (see Wakelam et al. 2010b for a detailed review on these processes). In addition to the reaction rate coefficients, the model predictions depend on other parameters such as the temperature, the density and the cosmic-ray ionization rate. Sensitivity analysis of the model predictions to the parameters by Wakelam et al. (2010a) showed that reaction rate coefficients are the most important model parameters and that the predicted abundances are strongly sensitive to their uncertainties. Current chemical models can include thousands of reactions for which only a small percentage has been studied in the extreme conditions of the cold (or even hot) ISM. For this reason, many of the rate coefficients are quite uncertain but the main problem is that we do not even know their real uncertainty. One way of quantitatively improving the model predictions is to identify, among the thousands of reactions, those which model abundances are mostly sensitive to. Those reactions can them be studied in details by experts and their rate coefficients estimated with a better precision.

Since the first chemical models for the ISM, astrochemists, in collaboration with physico-chemists, have improved over the years our knowledge of the micro-physics at play in these environments but the road is still long (Smith 2011, Larsson et al. 2012). With the release of a new, interactive, online database for chemical reactions for ISM simulations, the KInetic Database for Astrochemistry[1] (Wakelam et al. 2012), a new impulse has been given to this field by offering the astrochemists with the opportunity to actively participate to the feeding of reaction databases. KIDA is also a platform where physico-chemists can show their results.

Within the KIDA scientists, a large program on the nitrogen chemistry has been undertaken especially because of the importance of neutral-neutral reactions poorly known. Experimental measurements at low temperature of the atomic nitrogen reactivity with three radicals (NO, OH and CN), governing the $N_2$ formation, have shown that the formation of molecular nitrogen was much less efficient than previously assumed (Bergeat et al. 2009, Daranlot et al. 2011, Daranlot et al. 2012). In parallel, a theoretical review on about 20 reactions important for the nitrogen chemistry in cold, dense ISM environments has been undertaken and is presented in this paper. All these reactions were identified using sensitivity analysis by Wakelam et al. (2010b).

## 2. List of reactions and comments on the recommended rate coefficients

The list of studied reactions is given in Table 1. The reactants and products of the reactions are in the first column. All species are considered in the ground state (not excited). When excited species are produced (N + CH for example), the species are assumed to relax before they can react. The second and third columns of Table 1 give the reaction rate coefficient as a function of temperature and the uncertainties in those rate coefficients that we propose. Uncertainties are defined by two numbers : $F_0$ is the uncertainty factor, assuming that the distribution of the rate coefficient follows a lognormal distribution, and $g$ gives the temperature dependence of this uncertainty. Details on this formalism are given in Wakelam et al. (2012). The range of temperature over which our recommendation is valid is given in the fourth column and we report the rate coefficient in the OSU (09-2008) database in the last one.

The details of the review if given in the form of a datasheet for each reaction that have been added to the KIDA database. The format of the datasheets is similar to the one used by the IUPAC experts to give recommendations. Information on experimental or theoretical data in the literature is given and commented in the database. Based on this information, recommendations on the rate coefficients and uncertainties to be used over the mentioned range of temperature is given.

Among the reactions listed in Table 1, half of them were already discussed in Wakelam et al. (2010b) but some of them had their rate coefficient revised considering new studies and the others were included in this paper in order to publish the datasheets. Many of those rate coefficients were changed compared to the 2008 version of the OSU database and the kida.uva.2011 network for the dense ISM (Wakelam et al. 2012). The new rate coefficients have been included into KIDA (with the corresponding datasheet) and will be in the next version of the kida.uva database that will be released at the beginning of the year 2014. Although a datasheet for the reaction N + $C_3$ is available, the reaction has not been included in Table 1 because we could not give any law as a function of temperature. The rate coefficient at 10 K is smaller than $10^{-16}$ $cm^3$ $s^{-1}$ but no precise value is available. Considering its lack of importance at such small values, we simply recommend to ignore this reaction.

## 3. Conclusion

Key reactions for the nitrogen chemistry in the interstellar medium have been identified using sensitivity analyses in a previous study by Wakelam et al. (2010b). Some recommendations on the rate coefficients to use for some of these reactions had been proposed. We complete this study by presenting in this paper, recommendations for other reactions but also changing some of the rate coefficients based on more recent experimental and/or theoretical studies. The justification of those recommendations is given in online appendix datasheets for each reaction. We encourage astrophysicists to use those new values for any application within the recommended range of temperature.

[1] http://kida.obs.u-bordeaux1.fr/

Acknowledgements
This research has been partially funded by the CNRS/INSU (PCMI, PNP and ASOV) and the Observatoire Aquitain des Sciences de l'Univers. We acknowledge support from the Agence Nationale de la Recherche (ANR-JC08–311018: EMA:INC). The International Space Science Institute provided some of the authors with the opportunity to start this project in the context of the international team with the title "New generation of databases for interstellar chemical modeling in preparation for HSO and ALMA". VW and IWMS thank the Royal Society for its financial support of their collaboration. SJK acknowledges support through NASA - PATM grant number NNH09AK24I. We acknowledge support from the COST Action CM0805.

Table 1
List of studied reactions and associated rate coefficients

| Reaction | $k$(T) ($cm^3s^{-1}$) | Uncert. ($F_0$ ; $g$) | T range (K) | OSU |
|---|---|---|---|---|
| N + CH → CN + H | $1.4 \times 10^{-10}$ $(T/300)^{0.41}$ | 1.3 ; 4 | 56 - 300 | $1.66 \times 10^{-10}$ $(T/300)^{-0.09}$ |
| N + CN → $N_2$ + C | $8.8 \times 10^{-11}(T/300)^{0.42}$ | 1.4 ; 1.5 | 10 - 300 | $3.0 \times 10^{-10}$ |
| N + OH → H + NO | $5 \times 10^{-11}exp(-6/T)$ | 1.4 ; 7 | 10 - 500 | $7.5 \times 10^{-11}$ $(T/300)^{-0.18}$ |
| N + NO → $N_2$ + O | $4 \times 10^{-11}(T/300)^{-0.2}exp(-20/T)$ | 1.4 ; 10 | 10 - 500 | $3.0 \times 10^{-11}$ $(T/300)^{-0.6}$ |
| N + NH → $N_2$ + H | $5.0 \times 10^{-11}$ $(T/300)^{0.1}$ | 2 ; 6 | 10 - 500 | $5.0 \times 10^{-11}$ |
| N + $NH_2$ → $N_2$ + H + H | $1.2 \times 10^{-10}$ | 1.3 ; 6 | 10 - 500 | - |
| N + $C_2N$ → CN + CN | $1.0 \times 10^{-10}$ | 3 ; 2.97 | 10 - 300 | $1.0 \times 10^{-10}$ |
| N + $C_4N$ → CN + $C_3N$ | $9.0 \times 10^{-11}(T/300)^{0.17}$ | 3 ; 0 | 10 - 300 | $1.0 \times 10^{-10}$ |
| N + $CH_2$ → HCN + H<br>N + $CH_2$ → HNC + H | $5.0 \times 10^{-11}$ $(T/300)^{0.17}$<br>$3.00 \times 10^{-11}$ $(T/300)^{0.17}$ | 1.5 ; 0<br>1.5 ; 0 | 10 - 300<br>10 - 300 | $3.95 \times 10^{-11}$ $(T/300)^{0.17}$<br>$3.95 \times 10^{-11}$ $(T/300)^{0.17}$ |
| C + OCN → CN + CO | $1.0 \times 10^{-10}$ | 5 ; 0 | 10 - 300 | $1.0 \times 10^{-10}$ |
| C + $NH_2$ → HCN + H<br>C + $NH_2$ → HNC + H | $3.0 \times 10^{-11}$ $(T/300)^{-0.2}exp(-6/T)$<br>$3.0 \times 10^{-11}$ $(T/300)^{-0.2}exp(-6/T)$ | 1.5 ; 0<br>1.5 ; 0 | 10 - 300<br>10 - 300 | $3.4 \times 10^{-11}$ $(T/300)^{-0.36}$<br>$3.4 \times 10^{-11}$ $(T/300)^{-0.36}$ |
| O + $C_3N$ → CO + $C_2N$ | $1.0 \times 10^{-10}$ | 3 ; 2.97 | 10 - 300 | $4.0 \times 10^{-11}$ |
| O + CN → N + CO | $5.0 \times 10^{-11}$ | 3 ; 0 | 10 - 300 | $4.0 \times 10^{-11}$ |
| O + HNO → $NO_2$ + H<br>O + HNO → NO + OH | 0<br>$3.8 \times 10^{-11}$ $(T/300)^{-0.08}$ | -<br>2 ; 7 | 10 - 300<br>10 - 300 | $1.0 \times 10^{-12}$<br>$3.8 \times 10^{-11}$ |
| O + NH → NO + H<br>O + NH → OH + N | $6.6 \times 10^{-11}$<br>0 | 3 ; 2.97<br>- | 10 - 300<br>10 - 300 | $1.16 \times 10^{-10}$<br>$1.16 \times 10^{-11}$ |
| O + $NH_2$ → HNO + H<br>O + $NH_2$ → NO + $H_2$<br>O + $NH_2$ → OH + NH | $6.3 \times 10^{-11}$ $(T/300)^{-0.1}$<br>0<br>$7.0 \times 10^{-12}$ $(T/300)^{-0.1}$ | 2 ; 4<br>-<br>2 ; 4 | 10 - 300<br>10 - 300<br>10 - 300 | $8.0 \times 10^{-11}$<br>$1.0 \times 10^{-11}$<br>$7.0 \times 10^{-12}$ |
| CH + $NH_3$ → $H_2CNH$ + H<br>CH + $NH_3$ → HCN + $H_2$ + H | $1.52 \times 10^{-10}$ $(T/300)^{-0.05}$<br>$8.0 \times 10^{-12}$ $(T/300)^{-0.05}$ | 1.2 ; 2<br>1.2 ; 2 | 10 - 300<br>10 - 300 | -<br>- |

| Reaction | $k$(T) ($cm^3s^{-1}$) | Uncert. ($F_0$ ; $g$) | T range (K) | OSU |
|---|---|---|---|---|
| $CN + NH_3 \rightarrow HCN + NH_2$ | $2.8 \times 10^{-11} (T/300)^{-0.85}$ | 1.2 ; 1.6 | 10 - 300 | $1.38 \times 10^{-11} (T/300)^{-1.14}$ |
| $CN + NH_3 \rightarrow NCNH_2 + H$ | 0 | 1.2 ; 1.6 | 10 - 300 | $1.3 \times 10^{-11} (T/300)^{-1.11}$ |
| $NH_3^+ + H_2 \rightarrow NH_4^+ + H$ | $3.36 \times 10^{-14} \exp(35.7/T)$ | 1.5 ; 0 | 10 - 20 | $1.5 \times 10^{-14} (T/300)^{-1.50}$ |
| | $2.0 \times 10^{-13}$ | 1.5 ; 0 | 21 - 230 | $1.5 \times 10^{-14} (T/300)^{-1.50}$ |
| | $1.7 \times 10^{-11} \exp(-1044/T)$ | 1.5 ; 0 | 231 - 800 | $1.5 \times 10^{-14} (T/300)^{-1.50}$ |
| $N_2H^+ + e^- \rightarrow N_2 + H$ | $2.47 \times 10^{-7} (T/300)^{-0.84}$ | 1.6 ; 0 | 10 - 1000 | $9.0 \times 10^{-8} (T/300)^{-0.51}$ |
| $N_2H^+ + e^- \rightarrow NH + H$ | $1.30 \times 10^{-8} (T/300)^{-0.84}$ | 1.6 ; 0 | 10 - 1000 | $1.0 \times 10^{-8} (T/300)^{-0.51}$ |

**Authors:**

**Ian Smith (University of Cambridge, UK)**

**Jean-Christophe Loison (Univ. of Bordeaux – CNRS, Fr)**

$N(^4S) + CH(^2\Pi) \rightarrow CN(X^2\Sigma^+) + H(^2S)$ **(1)** $\Delta H_r^{298} = -414$ kJ mol$^{-1}$ *(Baulch* et al.*, 2005)*

$\rightarrow CN(A^2\Pi) + H(^2S)$ **(2)** $\Delta H_r^{298} = -303$ kJ mol$^{-1}$

## Rate Coefficient Data *k*

| *k* / cm$^3$ molecule$^{-1}$ s$^{-1}$ | *T* / K | Reference | Comments |
|---|---|---|---|
| *Rate Coefficient Measurements* | | | |
| $(2.1 \pm 0.5) \times 10^{-11}$ | 298 | Messing *et al.* (1981) | |
| $(1.66 \pm 0.12) \times 10^{-10}\,(T/298)^{(-0.09\pm0.2)}$ | 216 - 584 | Brownsword *et al.*, (1996) | |
| $(1.4 \pm 0.4) \times 10^{-10}\,(T/298)^{(0.41\pm0.05)}$ | 56-296 | Daranlot et al, (2013) | (3) |
| *Reviews and Evaluations* | | | |
| $1.66 \times 10^{-10}\,(T/300)^{(-0.09)}$ | 222 – 584 | UMIST database | |
| $1.66 \times 10^{-10}\,(T/300)^{(-0.09)}$ | | OSU database | |

## Comments

The reactants correlate with triplet and quintet states, the products with only triplet states. The potential energy surfaces for this reaction are discussed in detail by Rayez et al. (4). A $^3A'$ PES connects the reactants with ground state products (reaction (1a)) and a $^3A''$ PES connects the reactants with $CN(B^2\Pi)$ + H (reaction (1b)). Therefore, there is an electronic degeneracy factor of *ca.* 3/8. The reaction probably proceeds *via* energised intermediates (excited triplet states of HCN).

Calculations have been performed (5) using long-range transition state theory (6). The long-range interaction between reactants is assumed to arise from dispersion forces and dipole-induced dipole forces. Spin-orbit effects are ignored and the rotation of CH is treated quasi-classically. The calculated rate constants between 10 and 320 K fit: $k(T) = 1.87 \times 10^{-10}\,(T/298)^{0.18}$ cm$^3$ molecule$^{-1}$ s$^{-1}$. The results in the temperature range covered in the measurements in ref. (3) agree quite well with the experimental values.

The experiments of Brownsword et al. (2) should be reliable. The large values of the observed rate coefficients, and their agreement with the values calculated by long-range transitions state theory, suggest that the reaction rate is determined by capture under the influence of long-range forces. We recommend the calculated rate constant scaled to the experimental results at 298K.

## Preferred Values

*Between 56 and 300 K :*

$k(T) = 1.4 \times 10^{-10} \times (T/298)^{0.41}$ cm$^3$ molecule$^{-1}$ s$^{-1}$

*Reliability*

F = 1.3, g = 4.0

*Comments on Preferred Values*

The UMIST and OSU data bases adopt the rate coefficients measured by Brownsword et al. (2) – and their *T*-dependence. The reason for the 'low' values reported by Messing et al. (1) is not clear. We recommend the new rate constant obtained in CRESU experiment. For temperatures below 56 K, the rate constant is more uncertain so we introduced a large "g".

**Authors:**

**Ian Smith (University of Cambridge, UK)**

**Stephen Klippenstein (Argonne National Laboratory, Argonne, IL, USA)**

**Jean-Christophe Loison, Astrid Bergeat and Kevin M. Hickson (Univ. Bordeaux, France)**

**$N(^4S) + CN(^2\Sigma^+) \rightarrow C(^3P) + N_2(^1\Sigma_g^+)$** $\Delta H^o_{298}(1) = -191.4$ kJ mol$^{-1}$ *(Baulch* et al.*, 2005)*

## Rate Coefficient Data *k*

| $k$ / cm$^3$ molecule$^{-1}$ s$^{-1}$ | $T$ / K | Reference | Comments |
|---|---|---|---|
| *Rate Coefficient Measurements* | | | |
| $1.0 \times 10^{-10}$ | 300 | Whyte & Phillips, 1983 | |
| $3.24 \times 10^{-13}\exp(1770/T)$ | 300-534 | Atakan *et al.*, 1992 | |
| $8.8 \times 10^{-11}\,(T/300)^{0.42}$ | 56-296 | Daranlot *et al.*, 2012 | |
| *Reviews and Evaluations* | | | |
| $9.8 \times 10^{-10}\,T^{-0.40}$ | 300-3000 | Baulch *et al.*, 2005 (p. 1139) | |
| $3.0 \times 10^{-10}$ | 298-2500 | UMIST database | |
| $3.0 \times 10^{-10}$ | all temperatures | OSU website | |
| *Theory* | | | |
| | 300-2500 | Moskaleva & Lin, 2001 | |
| $2.0 \times 10^{-10}\,(T/300)^{0.18}$ | 10-400 | Klippenstein & Harding, 2011 | |
| | 50-300 | Ma *et al.*, 2012 | |

## Comments

The reactants correlate with triplet and quintet states, the products with only triplet states. As the reactants are only in S and Σ states there is no spin orbit coupling. Therefore, there is a constant electronic degeneracy factor of *ca.* 3/8. The three measurements of the rate coefficient at 298 K agree well. However, Atakan *et al* (2) suggests what seems like an extraordinarily steep negative dependence with *T*. L. B. Harding performed (for this datasheet) CASPT2(10e,9o)/CBS scans of the potential energy surface for this reaction. These calculations suggest that there is no barrier to formation of either NCN (exothermic by 435 kJ/mol) or CNN (also exothermic by 337 kJ/mol). Furthemore, the saddlepoint for transformation from NCN to CNN is well below the N + CN energy. So the reaction likely proceeds via addition to form both NCN and CNN followed by dissociation from the CNN complex to C + NN. However, even though very exothermic, either the isomerization or the dissociation could provide some sort of dynamical bottleneck especially at higher *T*.

S. Klippenstein performed long-range TST calculations (using CASPT2(8,8)/CBS potentials and so including not only dispersion but also contributions from the dipole induced-dipole and other terms) leading to (including the 3/8 term from electronic degeneracy) $k$(CN+N) = $2.0 \times 10^{-10}\,(T/300)^{1/6}$ cm$^3$ molecule$^{-1}$ s$^{-1}$, which is about twice the experimental determinations (1 and $1.1 \times 10^{-10}$ cm$^3$ molecule$^{-1}$ s$^{-1}$) at room temperature. The CRESU data by Daranlot *et al.* (3) obtained at low temperature present a more pronounced positive temperature dependence. The rate constants were determined relative to those of the N+OH reaction. Ma *et al.* (5) have performed quantum capture calculations on a new two-dimensional potential energy surface to calculate low-temperature rate constants for the N + CN reaction. These rate constants present a positive temperature dependence in reasonably good agreement with the experimentally determined relative rate values of Daranlot *et al.* (3).

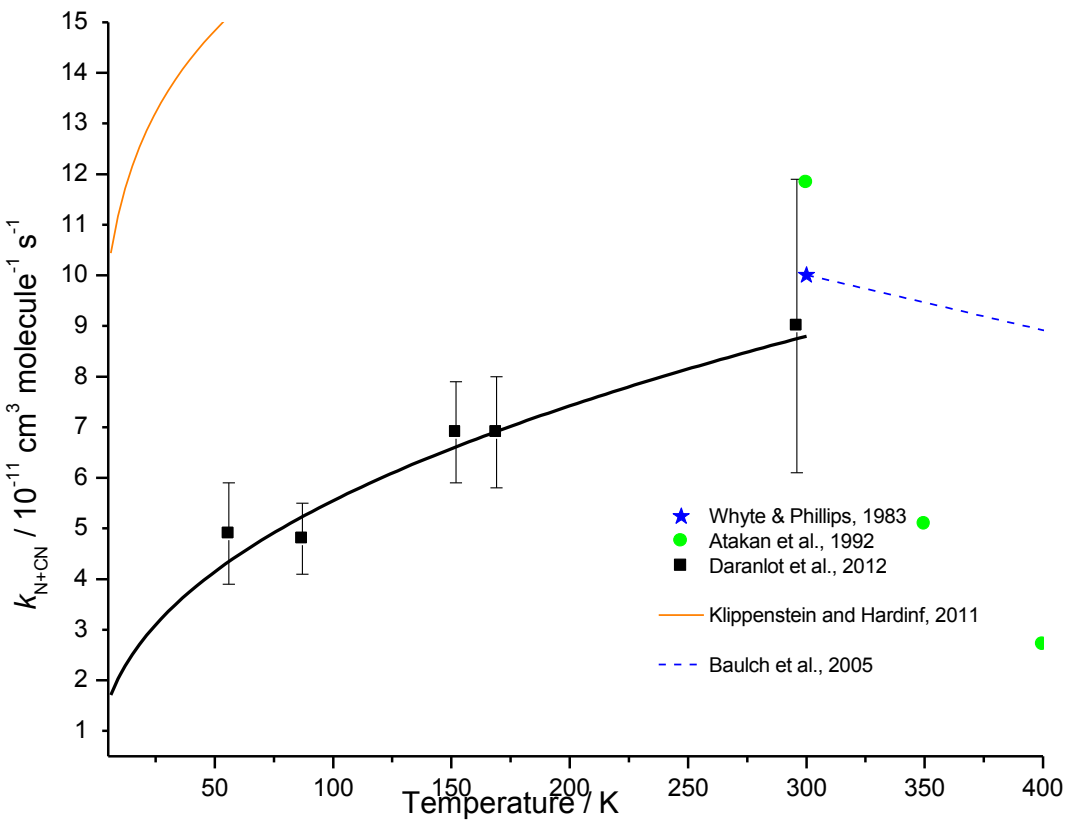

## Preferred Values

*Rate coefficient (10 – 300 K)*

$\boldsymbol{k}$ **(T) = 8.8×10$^{-11}$ (T/300)$^{0.42}$ cm$^3$ molecule$^{-1}$ s$^{-1}$**

*Reliability*

**$F_0$ = 1.4 ; g = 1.5**

**Authors:**
**Ian Smith (University of Cambridge, UK)**
**Jean-Christophe Loison (Univ. of Bordeaux - CNRS, Fr)**
**Stephen Klippenstein (Argonne National Laboratory, Argonne, IL, USA)**

**$N(^4S) + C_3(^1\Sigma_g^+) \rightarrow CN(^2\Sigma^+) + C_2(X^1\Sigma_g^+)$ (1)** $\Delta H_r^{298} = -$ 20 kJ/mol *(Baulch* et al.*, 2005)*

**$\rightarrow CN(^2\Sigma^+) + C_2(a^3\Pi)$ (2)** $\Delta H_r^{298} = -$ 13 kJ/mol *(Baulch* et al.*, 2005)*

## Rate Coefficient Data ($k = k_1 + k_2$)

| $k$ / cm$^3$ molecule$^{-1}$ s$^{-1}$ | $T$ / K | Reference | Comments |
|---|---|---|---|
| *Rate Coefficient Measurements (k)* | | | |
| No records in the NIST database | | | |
| *Reviews and Evaluation* | | | |
| $k = 1.0 \times 10^{-13}$ | 10 – 300 | UMIST database | |
| $k = 1.0 \times 10^{-13}$ | no $T$-dependence | OSU website | |

## Comments

This radical-radical reaction is slightly exothermic. Reaction to the ground ($^1\Sigma_g^+$) state of $C_2$ is spin-forbidden. However, reaction to $C_2(^3\Pi)$ is exothermic and spin-allowed. On the other hand, $N(^4S)$ atoms are generally not reactive to other species in singlet states and $C_3$ is also not very reactive, even to other radicals such as NO and $O_2$.

If we consider $C_3$ to be an 'honorary' unsaturated hydrocarbon, then by the 'rules' proposed by (1) this reaction would be slow (N atoms have a negative electron affinity). To confirm this conclusion we performed *ab initio* calculations at various levels: (CCSD(T)/CBS//CCSD(T)/cc-pVQZ (T1 diagnostic ~0.03) leading to a barrier equal to 12 kJ/mol, CASPT2(11e,11o) leading to a barrier equal to 7 kJ/mol, MRCI+Q(16,8)/VTZ leading to a barrier equal to 18 kJ/mol and finally UHF/DFT (M06-2X/vtz) leading to an absence of barrier. These results strongly suggest that there is a barrier of a few kJ/mol preventing the addition of $N(^4S)$ to $C_3(^1\Sigma_g^+)$ at low temperature, even if for Titan's atmosphere (150-170 K) it is less obvious that this reaction can be neglected.

It seems that this reaction is probably slow at 298 K and very slow indeed at 10 K. The origin of the values of $k(T)$ in the UMIST and OSU data bases is unknown.

## Preferred Values

### *Rate coefficients (10 – 300 K)*

$k_1$(300 K) = $k_1$(10 K) = 0
$k_2$(300 K) = 1 x $10^{-13}$ cm$^3$ molecule$^{-1}$ s$^{-1}$
$k_2$(10 K) < 1 x $10^{-16}$ cm$^3$ molecule$^{-1}$ s$^{-1}$

*Reliability*
$\mathbf{F_0 = 10}$ **;** $\mathbf{g = 0}$

### *Comments on Preferred Values*

The rate coefficient at 298 K of $10^{-13}$ cm$^3$ s$^{-1}$ is the best estimate. At 10 K, it seems safe to assume that the reaction has a negligible rate. No temperature dependence rate coefficient can be given in the absence of more detailed calculations.

Note that in the interstellar medium, electronically excited CN will relax before reacting, for this reason, in most astrophysical environments, we do not make any distinctions between ground and excited states.

**Authors:**

**Astrid Bergeat and Kevin M. Hickson (Univ. Bordeaux, France)**

$$N(^4S) + OH(X^2\Pi) \rightarrow H(^2S) + NO(X^2\Pi) \quad \textbf{(1)}\ \Delta H^o_{298}(1) = -201.59\ \text{kJ mol}^{-1}$$

$$\rightarrow O(^3P) + NH(X^3\Sigma^-) \quad \textbf{(2)}\ \Delta H^o_{298}(2) = +95.8\ \text{kJ mol}^{-1}$$

*Thermodynamic Data from* (Baulch *et al.*, 2005)

**Rate Coefficient Data *k***

| $k$ / $cm^3$ molecule$^{-1}$ $s^{-1}$ | $T$ / K | Reference | Comments |
|---|---|---|---|
| *Rate Coefficient Measurements* | | | |
| | 50 – 294 | Daranlot *et al.*, 2011 | [2] |
| $k_1 = 2.0\times10^{-10}$ $(T/298)^{-0.17}\times3/[4*(2+\exp(-205/T))]$ | 103 – 294 | Smith and Stewart, 1994 | [3] |
| $k_1 = (4.2\pm0.8)\times10^{-11}$ | 298 | Brune *et al.*, 1983 | [4] |
| $k_1 = (2.21\pm0.18)\times10^{-10}$ $T^{-0.25\pm0.17}$ | 250 – 515 | Howard and Smith, 1981 | [5] |
| *Theory* | | | |
| | 5 – 500 | Daranlot *et al.*, 2012 | [6] |
| | 5 – 500 | Li *et al.*, 2011 | [7] |
| | 5 – 500 | Jorfi *et al.*, 2009 | [8] |
| | 5 – 500 | Ge *et al.*, 2008 | [9] |
| | 5 – 515 | Edvardsson *et al.*, 2006 | [10] |
| | 300 – 500 | Chen *et al.*, 2003 | [11] |
| $8.41\times10^{-12}$ $T^{0.30}$ | 5 – 200 | Cobos, 1995 | [12] |
| *Reviews and Evaluations* | | | |
| $k_1 = 1.8\times10^{-10}$ $T^{-0.2}$ | 100-2500 | Baulch *et al.* 2005 | [1] |
| *Preferred value* | | | |
| $4.5\times10^{-11}$ | 100 – 500 | | |

## Comments

[1] Evaluation of literature data up to 1994. Recommendation mainly based on the two experimental temperature dependences.

[2] Experiments in a continuous supersonic flow reactor. N atoms were produced by microwave discharge upstream of the Laval nozzle OH radicals produced by pulsed laser photolysis of $H_2O_2$ and probed by laser-induced fluorescence.

[3] OH radicals produced by pulsed laser photolysis of $HNO_3$ and probed by laser-induced fluorescence. N atoms produced by microwave discharge and titrated by NO. Cryogenically cooled flow tube. Errors quoted as single standard deviations.

[4] OH radicals produced by the reaction F + $H_2O$ and N atoms produced by microwave discharge. Radical concentrations were determined by a variety of techniques and titration: laser magnetic resonance, resonance fluorescence and resonance absorption.

[5] OH radicals produced by flash photolysis of $H_2O$ and probed by resonance fluorescence with a lamp. N atoms produced by microwave discharge and titrated by NO in Ar. Cooled or heated flow tube. Cited uncertainties are 95% confidence limits.

[6] Calculations: (1) time-independent quantum mechanical (TIQM) method and J-shifting approach and (2) time-dependent quantum method (TDQM) including contributions from all angular momenta $J$ on the high-quality potential energy surface (PES) of the $a^3A''$ state

(the lowest triplet electronic state of HNO) of Li *et al*. 2011.
[7] PES and all *J* TDQM calculations down to 100 K.
[8] Quasi-classical trajectory (QCT) calculations on an *ab initio* global PES $X^3A''$ of Guadagnini *et al*. 1995.
[9] TDQ wave packet (WP) method, on the same PES (Guadagnini *et al.* 1995) under both coupled-state or centrifugal sudden (CS) approximation and Coriolis-coupled or close-coupling (CC) approach.
[10] Calculations using the rotationally adiabatic capture centrifugal sudden approximation (ACCSA) in combination with *ab initio* electronic structure theory. For 103 K $\leq T$, $k_r = (4.03\pm0.02) \times 10^{-11} \exp((25\pm1)/T)$ $cm^3$ $molecule^{-1}$ $s^{-1}$.
[11] Three-dimensional TDQWP calculations on the potential energy surface of Guadagnini *et al*. 1995.
[12] Statistical adiabatic channel model (SACM) on a PES based on ab initio quantum chemical data of Pauzat *et al.* 1993.

This atom-radical reaction has been studied up to 3000 K.

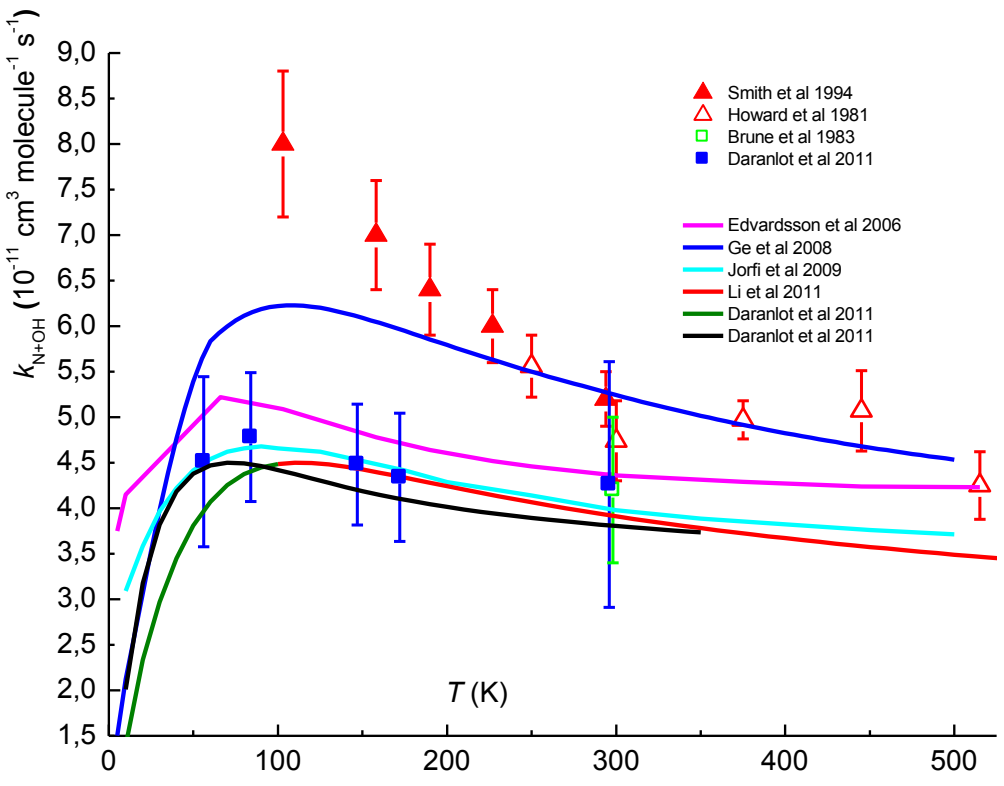

## Preferred Values

*Rate coefficient (10 – 300 K)*
$\boldsymbol{k}$ **(T) = $5\times10^{-11}$ exp(-6/T) $cm^3$ $molecule^{-1}$ $s^{-1}$**

*Reliability*
**$F_0$ = 1.4 ; g = 7**

*Comments on Preferred Values*

**Authors:**

**Ian Smith (University of Cambridge, UK)**

**Astrid Bergeat and Kevin M. Hickson (Univ. Bordeaux, France)**

## $N(^4S) + NO(X^2\Pi) \rightarrow N_2(X^1\Sigma^+_g) + O(^3P)$ $\Delta H^\circ_{298}(1) = -313.8$ kJ mol$^{-1}$ *(Baulch* et al.*, 2005)*

**Rate Coefficient Data *k***

| $k$ / cm$^3$ molecule$^{-1}$ s$^{-1}$ | $T$ / K | Reference | Comments |
|---|---|---|---|
| *Rate Coefficient Measurements* | | | |
| $(2.8 \pm 0.1)\times10^{-11}(T / 300)^{-0.40 \pm 0.15}$ | 196 – 370 | Lee *et al.*, 1978 | [3] |
| $(2.2 \pm 0.2)\times10^{-11} \exp\{(160 \pm 50) / T\}$ | 213 – 369 | Wennberg *et al.*, 1994 | [4] |
| $(3.2 \pm 0.6)\times10^{-11} \exp\{(25 \pm 16) / T\}$ | 48 – 211 | Bergeat *et al.*, 2009 | [5] |
| *Theory* | | | |
| $3.4\times10^{-11} \exp(-24.8 / T)$ | 100 – 1000 | Duff *et al.*, 1996 | [6] |
| | 10 – 500 | Jorfi *et al.*, 2009 | [7] |
| | 10 – 5000 | Gamallo *et al.*, 2010 | [8] |
| *Reviews and Evaluations* | | | |
| $3.5 \times 10^{-11}$ | 210 – 3700 | Baulch *et al.*, 2005 | [1] |
| $2.1\times10^{-11} \exp(+100 / T)$ | all temperatures | JPL Publication, 2006 | [2] |
| $3.75\times10^{-11} \exp(-26 / T)$ | 100 – 4000 | UMIST database, 2006 | |
| $3.0\times10^{-11} (T / 300)^{0.6}$ | all temperatures | OSU website, 2008 | |

## Comments

There are numerous studies at room temperatures and at higher temperatures (up to 3660 K).

[1] Evaluation of literature data up to 1996. Recommendation mainly based on the temperature independence at high temperatures and the small one at low temperature, which is consistent within the fairly substantial scatter in the measurements.

[2] Based on the temperature dependence of Wennberg *et al.* 1994 and measurements of Lee *et al.* 1978

[3] Discharge flow-filtered resonance fluorescence (DF-RF) and flash photolysis-resonance fluorescence. However, $N_2O$ photolysis is known to produce excited N atoms. The data obtained with this technique are thus not reported here and the other data (DF-RF) were corrected for axial diffusion and fitted by weighted least squares analysis.

[4] N atoms produced by microwave discharge of trace $N_2$ in He and monitored by atomic resonance fluorescence using a gas filter scheme.

[5] Experiments in a continuous supersonic flow reactor. N atoms were produced by microwave discharge upstream of the Laval nozzle and were probed in the vacuum ultraviolet by resonance fluorescence.

[6] Quasiclassical trajectory calculations on the $^3$A" surface of Walch *et al.* 1987 which presents a small energy barrier (in the uncertainty). The temperature dependence of the spin-orbit coupling effect was not taken into account.

[7] Time-independent quantum mechanical calculation with *J*-shifting approximation. Potential energy surfaces (PES) from Gamallo *et al.* 2010

[8] Time-dependent real wave-packet (WP) quantum dynamics rate constants on the 1 $^3$A" and 1 $^3$A' analytical PES and quasiclassical trajectory (QCT) method. The $^3$A" PES is barrierless along the minimum energy path, while the analytical $^3$A' excited PES presents an energy barrier of 36.57 kJ mol$^{-1}$, including zero point energy. The WP rate constant values

are in good agreement with the laboratory values.

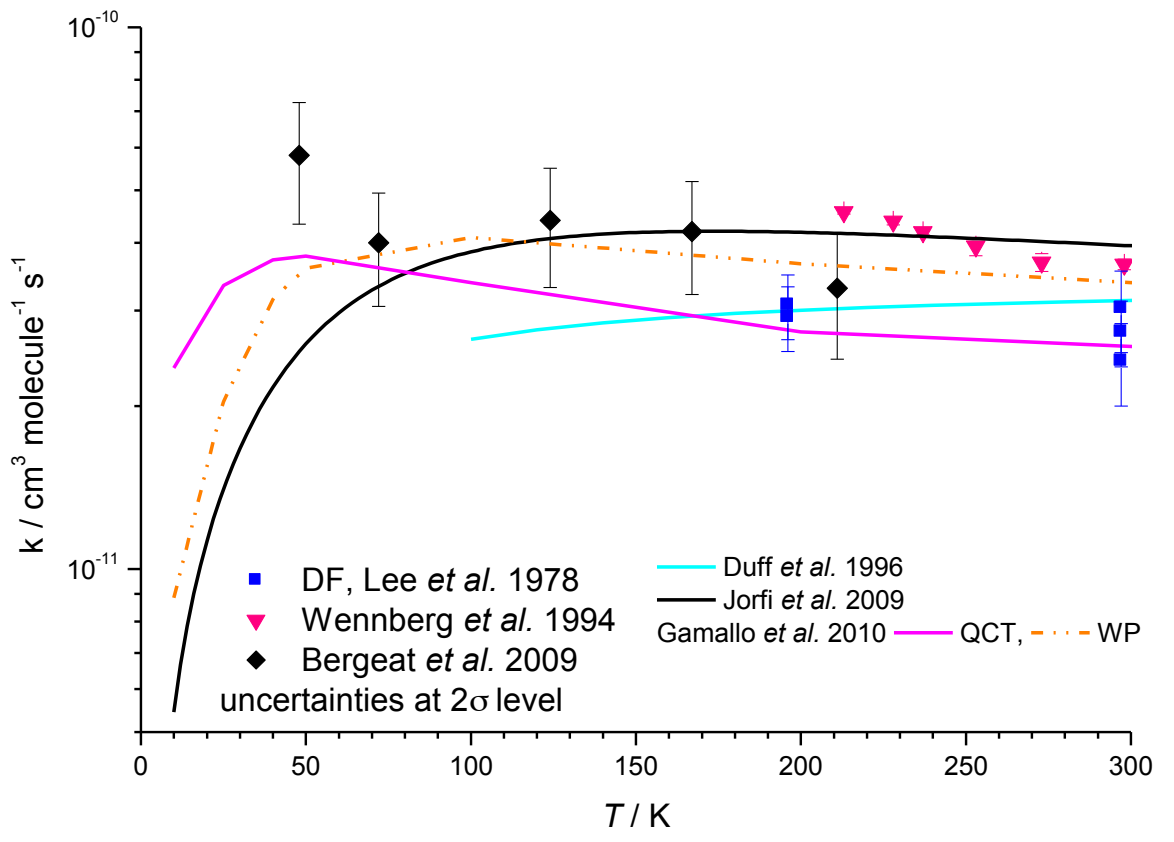

## Preferred Values

*Rate coefficient (10 – 300 K)*

$\boldsymbol{k}$ **(T) = 4×10$^{-11}$ (T/300)$^{-0.2}$ exp(-20/T) cm$^3$ molecule$^{-1}$ s$^{-1}$**

*Reliability*

**$F_0$ = 1.4 ; g = 10**

*Comments on Preferred Values*

**Author:**

**Jean-Christophe Loison (Univ. of Bordeaux - CNRS, France)**

## $N(^4S) + NH(X^3\Sigma^-) \rightarrow N_2(X^1\Sigma^+) + H(^2S)$ $\Delta Hr^{298} = -613$ kJ mol$^{-1}$ *(Baulch* et al*., 2005)*

## Rate Coefficient Data *k*

| *k* / cm$^3$ molecule$^{-1}$ s$^{-1}$ | *T* / K | Reference | |
|---|---|---|---|
| *Rate Coefficient Measurements* | | | |
| $2.49 \times 10^{-11}$ | 298 | Hack *et al*, 1994 | (1) |
| *Theory* | | | |
| $1.95\times10^{-11}\times (T/300)^{-0.51} \times \exp(-6.3/T)$ | 300-3000 | Caridade et al, 2005 | (2) |
| $7.3\times10^{-11}\times (T/300)^{0.094}$ | 100-3000 | Caridade et al, 2007 | (3) |
| $7.6\times10^{-11}\times (T/300)^{0.116}\times \exp(-0.792/T)$ | 2-300 | Frankcombe and Nyman (2007) | (4) |
| *Reviews and Evaluation* | | | |
| this reaction is not included | | UMIST database | |
| this reaction is not included | | OSU website | |

## Comments

The $N(^4S) + NH(X^3\Sigma^-)$ correlate with sextuplet, quadruplet and doublet states and $N_2(X^1\Sigma^+) + H(^2S)$ only with doublet states. Then, there is a 2/(2+4+6)=1/6 degeneracy factor. There has been only one direct experimental investigation of the rate coefficient for this reaction at room temperature.(1) The reaction is found to be relatively rapid at room temperature *(k* = $2.49 \times 10^{-11}$ cm$^3$ molecule$^{-1}$ s$^{-1}$). Ab-initio calculations have been performed on this system showing no barrier for $^2$N-NH formation. (3) Along the minimum energy path the reaction proceeds to a strongly bound collision complex ($N_2H$) without a potential barrier. A very small barrier exists on the exit channel from $N_2H$ to the $N_2$ + H products, well below the energy of the entrance channel. The overall reaction is exothermic by 613 kJ/mol (6.33 eV), implying a very strong tendency for any $N_2H$ formed to go on to produce the $N_2$ + H products. Various kinetic calculations have been performed on this system such as quasi-classical trajectory (QCT)(2,3), capture theory(3) and adiabatic capture, centrifugal sudden approximation (ACCSA).(4) It should be noted that dynamics calculations of Caridade et al in 2005(2) were affected by an error in the collision energy sampling, favoring high-energy values.

The results summarized on the figure below (from Frankcombe and Nyman (4)):

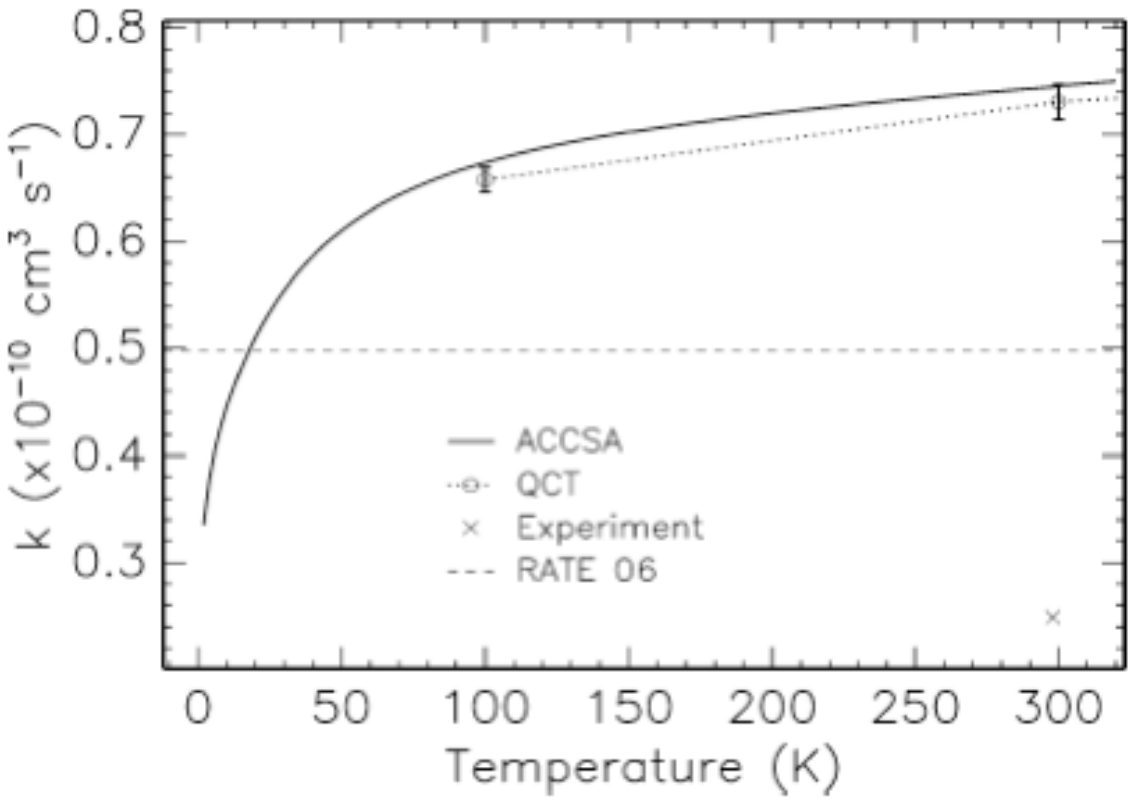

There is clear disagreement between the calculated rates constant and the single existing experimental measurement. The agreement between the QCT and ACCSA calculations indicate that the calculated rate coefficients represent the true rate coefficients on the potential energy surface, itself derived from high quality ab initio data. The difference with the experimental measurements may be due to the fact that the experimental rate coefficient have been determined by modeling NH reactant concentrations profiles as a function of the reaction time, with and without N atoms and then is very sensitive to the atomic concentrations. However, as noted by Caridade et al(3), the partition of the total rate of NH removal may not be equal for both doublet electronic states. Thus, the observed discrepancies between the calculated and measured rate coefficients may be attributed both to experimental difficulties and to the non-inclusion of non-adiabatic effects in the theory as complicated electronic crossings.

We recommend the use of the T dependency obtained by making an average of the very similar theoretical T dependency, with k(300K) value being the average of theoretical and experimental values.

## Preferred Values

*Rate coefficient (10 – 500 K)*

$\boldsymbol{k}$ **(T) = 5×10$^{-11}$ (T/300)$^{0.1}$ cm$^{3}$ molecule$^{-1}$ s$^{-1}$**

*Reliability*

**$F_0$ = 2 ; g = 6**

*Comments on Preferred Values*

**Author:**

**Jean-Christophe Loison (Univ. of Bordeaux - CNRS, Fr)**

## $N(^4S) + NH_2(X^2B_1) \rightarrow N_2(X^1\Sigma^+) + H(^2S) + H(^2S)$ $\Delta H_r^{298} = -227$ kJ mol$^{-1}$ *(Baulch* et al.*, 2005)*

## Rate Coefficient Data *k*

| *k* / cm$^3$ molecule$^{-1}$ s$^{-1}$ | *T* / K | Reference | Comments |
|---|---|---|---|
| *Rate Coefficient Measurements (k)* | | | |
| (1.21± 0.14) × 10$^{-10}$ | 296 | Whyte and Phillips, 1983 | (1) |
| (1.15± 0.21) × 10$^{-10}$ | 298 | Dransfeld and Wagner, 1987 | (2) |
| Mechanistic study | 296 | Whyte and Phillips, 1984 | (3) |
| *Reviews and Evaluation* | | | |
| this reaction is not included | | UMIST database | |
| this reaction is not included | | OSU website | |

## Comments

There have been two experimental investigations of the rate coefficient for this reaction[1-2] at room temperature, both giving similar results. The reaction is found to be rapid at room temperature *(k* = 1.2 × 10$^{-10}$ cm$^3$ molecule$^{-1}$ s$^{-1}$), so it has no barrier and should have a high rate at low temperature. This reaction has two exothermic product channels $N_2 + H_2$ or $N_2$ + H + H. The $N_2 + H_2$ channel is spin forbidden. Whyte and Phillips[3] performed a H atom branching ratio measurement for this reaction showing that the exit channel is the $N_2$ + H + H one. The absence of a barrier in the entrance channel means that the reaction is driven by long range interactions, mainly through dispersion. The high value of the rate coefficient at room temperature shows that there is no submerged barrier and the long range interaction term will lead to no substantial temperature dependence. We assume a constant value of the rate coefficient the 10-500 K range, the endothermic $NH(X^3\Sigma^-) + NH(X^3\Sigma^-)$ channel playing eventually a role only at even higher temperature.

## Preferred Values

*Rate coefficient (10 – 500 K)*

***k* (T) = 1.2×10$^{-10}$ cm$^3$ molecule$^{-1}$ s$^{-1}$**

*Reliability*

**$F_0$ = 1.3, g = 6**

*Comments on Preferred Values*

**Author:**

**Ian Smith (University of Cambridge, UK)**

## $N(^4S) + C_2N(X^2\Pi) \rightarrow CN(X\,^2\Sigma^+) + CN(X\,^2\Sigma^+)$ $\Delta Hr^{298} = -277$ kJ mol$^{-1}$ *(Baulch* et al.*, 2005)*

## Rate Coefficient Data *k*

| $k$ / cm$^3$ molecule$^{-1}$ s$^{-1}$ | $T$ / K | Reference | Comments |
|---|---|---|---|
| *Rate Coefficient Measurements* | | | |
| $1.0 \times 10^{-10}$ | 300 K | (a) | use the value given to model the time-dependence of CN in their experiments to find $k$ for N + CN. It seems a reasonable value but is not a *measurement.* I doubt that their fitting is sensitive to the value assumed. |
| *Reviews and Evaluations* | | | |
| $1.0 \times 10^{-10}$ | 10 - 300 | UMIST database | |
| $1.0 \times 10^{-10}$ | no $T$-dependence | OSU website | |

## Comments

This radical-radical reaction is strongly exothermic. Reaction is spin-allowed (over triplet PESs). However, the reactants also correlate with quintet PESs.

## Preferred Values

*Rate coefficient (10 – 300 K)*

$k$(300 K) = $1 \times 10^{-10}$ cm$^3$ molecule$^{-1}$ s$^{-1}$
$k$(10 K) = $1 \times 10^{-10}$ cm$^3$ molecule$^{-1}$ s$^{-1}$
**$k$(T) = $1 \times 10^{-10}$ cm$^3$ molecule$^{-1}$ s$^{-1}$**

*Reliability*

$\Delta \log k$ (300 K) = ± 0.5
$\Delta \log k$ (10 K) = ± 0.6
**$F_0 = 3$ ; g = 2.97**

*Comments on Preferred Values*

The UMIST and Ohio databases adopt the value given in (a). Despite my reservations about this experiment, I believe that the value they give is a reasonable estimate.

**Author:**

**Ian Smith (University of Cambridge, UK)**

**$N(^4S) + C_4N(^2\Sigma_g^+)$ → $CN(^2\Sigma^+) + C_3N(X^2\Sigma^+)$ (1)** No thermochemical data for $C_4N$ or $C_3N$

## Rate Coefficient Data *k*

| $k$ / cm$^3$ molecule$^{-1}$ s$^{-1}$ | $T$ / K | Reference | Comments |
|---|---|---|---|
| *Rate Coefficient Measurements (k)* | | | |
| No measurements were found in the literature | | | |
| *Reviews and Evaluations* | | | |
| $1.0 \times 10^{-10}$ | 10- 300 | UMIST database | |
| $1.0 \times 10^{-10}$ | no $T$-dependence | OSU website | |

## Comments

This radical-radical reaction is presumably exothermic (but I can find no thermochemical data on $C_4N$. Reaction is spin-allowed (over triplet PESs). However, the reactants also correlate with quintet PESs.

## Preferred Values

*Rate coefficient (10 – 300 K)*
**$k(T) = 9.0 \times 10^{-11}(T/300)^{0.17}$ cm$^3$s$^{-1}$**

*Reliability*
**$F_0 = 3$ ; $g = 0$**

*Comments on Preferred Values*
The UMIST and Ohio databases adopt the same rate coefficient value as for N + $C_2N$. This seems to be a reasonable estimate.

## References

**Authors:**

**Dahbia Talbi (Univ. of Montpellier – CNRS, Fr)**

**Jean-Christophe LOISON (Univ. of Bordeaux – CNRS, Fr)**

**$N(^4S) + CH_2(^3B_1)$ → $HCN(X\ ^1\Sigma^+) + H(^2S)$ (1)** $\Delta Hr_{298} = -510$ kJ/mol (Baulch *et al.* 2005)

**→ $HNC(X\ ^1\Sigma^+) + H(^2S)$ (2)** $\Delta Hr_{298} = -455$ kJ/mol (Baulch et al. 2005)

## Rate Coefficient Data ($k = k_1 + k_2$)

| $k$ / cm$^3$ molecule$^{-1}$ s$^{-1}$ | $T$ / K | Reference | Comments |
|---|---|---|---|
| *Rate Coefficient Measurements* | | | |
| No data was found | | | |
| *Calculations* | | | |
| $k(T) = 7.89 \times 10^{-11}(T/300)^{1/6}$ cm$^3$s$^{-1}$ | 10-300 | Herbst et al. (2000) | |
| *Reviews and Evaluations* | | | |
| $k_1(T) = 3.95 \times 10^{-11} (T/300)^{0.17}$ | 10-300 | UMIST database | |
| $k_2(T) = 3.95 \times 10^{-11} (T/300)^{0.17}$ | 10-300 | UMIST database | |
| $k_1(T) = 3.95 \times 10^{-11} (T/300)^{0.167}$ | | OSU database | |
| $k_2(T) = 3.95 \times 10^{-11} (T/300)^{0.167}$ | | OSU database | |

## Comments

There is no barrier for this reaction as shown by ab-initio calculations (Talbi 1999, Herbst *et al.* 2000, Takahashi & Takayanagi 2007). The ground state $^4N$ nitrogen atom has only one electronic state and $^4N$ + $^3CH_2$ correlate adiabatically with sextet, quadruplet and doublet state and products in their ground state correlate only to doublet state so there is an electronic degeneracy equal to 1/6 leading to capture rate constant equal to $8.0\times10^{-11}\times(T/300)^{0.17}$ cm$^3$ molecule$^{-1}$ s$^{-1}$ (Herbst et al. 2000). It should be noted that the quadruplet surface shows also no barrier in the entrance valley and may participate to the reaction, then the rate constant may be substantially higher, up to $2.4\times10^{-10}\times(T/300)^{0.17}$ cm$^3$ molecule$^{-1}$ s$^{-1}$ (Herbst et al. 2000). The main products from ab-intio results are the HCN + H. As the isomerization energy of the HCN → HNC is 186 kJ/mol (DePrince III & Mazziotti 2008), there is so much excess energy available in this reaction (510 kJ/mol) that the HCN product is able to undergo efficient isomerization reactions after production leading to near equal production rates of the two isomers. Precise internal energy distribution is necessary to get accurate branching ratio, the HCN being likely more abundant that HNC as all the HCN produced with less than 186 kJ/mol will be not able to isomerize. The final branching ratio will likely in favor of HCN, we propose a ratio HCN/HNC=5/3.

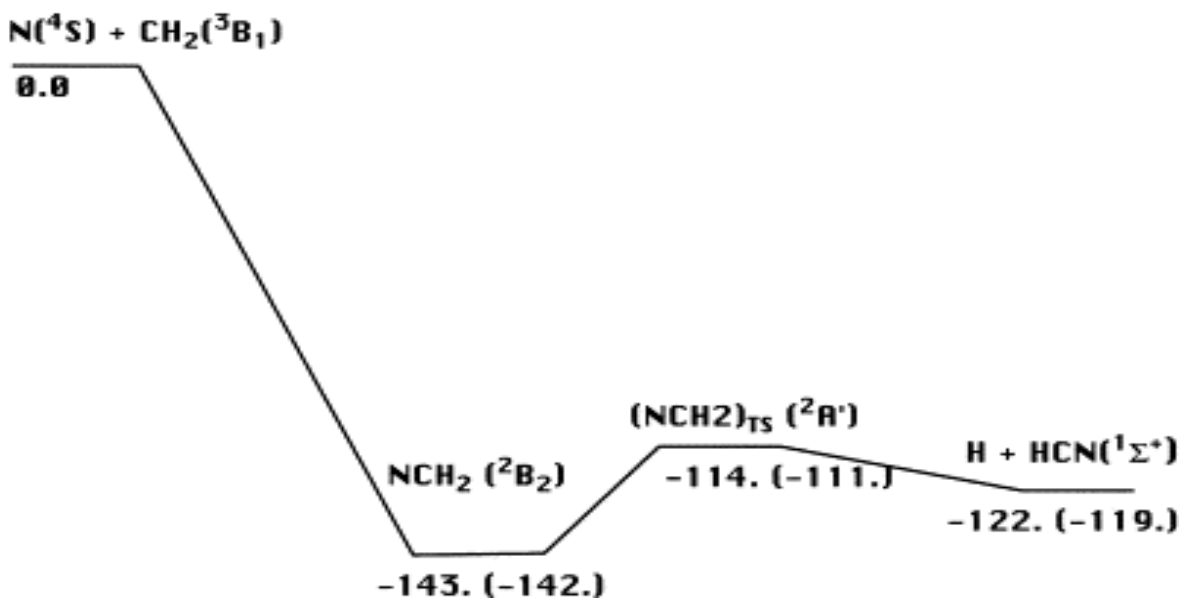

The Energy profiles are for surfaces of doublet multiplicity. Relative energies, given in kcal/mol, have been calculated at the PMP4SDTQ/6-311++G(3df,3pd)//MP2/6-31G(d,p) and CCSD(T)/6-311++G(3df,3pd)//MP2/6-31G(d,p) (numbers in brackets) levels. In all cases, relative energies are corrected for the ZPE and for spin contamination from higher spin state

### Preferred Values

*Rate coefficient (10 – 300 K)*

$k_1$ **(T) = 5.0×10$^{-11}$(T/300)$^{0.17}$ cm$^3$ s$^{-1}$**

$k_2$ **(T) = 3.0×10$^{-11}$(T/300)$^{0.17}$ cm$^3$ s$^{-1}$**

## Reliability

*$F_0$ = 1.5 ; g = 0*

## Comments on Preferred Values

**Author:**

**Ian Smith (University of Cambridge, UK)**

## $C(^3P) + OCN(^2\Pi) \rightarrow CN(^2\Sigma^+) + CO(^1\Sigma^+)$ (1) $\Delta Hr^{298} = -519$ kJ mol$^{-1}$ *(Baulch* et al.*, 2005)*

## Rate Coefficient Data *k*

| $k$ / cm$^3$ molecule$^{-1}$ s$^{-1}$ | $T$ / K | Reference | Comments |
|---|---|---|---|
| *Rate Coefficient Measurements* | | | |
| None that can be found. | | | |
| *Reviews and Evaluations* | | | |
| $1.0 \times 10^{-10}$ | 10 - 300 | UMIST database | |
| $1.0 \times 10^{-10}$ | $T$-independent | OSU website | |

## Comments

This is a very exothermic radical-radical reaction. The reactants correlate with 36 states ($3^2A' + 3^2A'' + 3^4A' + 3^2A''$), the products in their ground states with only one doublet surface. However the $A^2\Pi$ state of CN is energetically accessible and probably provides other adiabatic routes. Nevertheless, it seems that an electronic degeneracy factor of about 1/6 will decrease the rate coefficient from the simple collision value.

There have been no experiments on this reaction (or none listed in the NIST data base), and no calculations that I can find reference to.

## Preferred Values

*Rate coefficients (10 – 300 K)*

$k(300\text{ K}) = 1 \times 10^{-10}$ cm$^3$ molecule$^{-1}$ s$^{-1}$

$k(10\text{ K}) = 1 \times 10^{-10}$ cm$^3$ molecule$^{-1}$ s$^{-1}$

$\mathbf{k(T) = 1 \times 10^{-10}}$ **cm$^3$ molecule$^{-1}$ s$^{-1}$**

*Reliability*

$\Delta \log k$ (300 K) $= \pm 0.7$

$\Delta \log k$ (10 K) $= \pm 0.7$

**$F_0 = 5$ ; $g = 0$**

*Comments on Preferred Values*

In the absence of any data, the estimates given in the OSU and UMIST data bases appear reasonable for this radical-radical reaction. 'Reliability is low'.

Note that in the interstellar medium, electronically excited CN will relax before reacting, for this reason, in most astrophysical environments, we do not make any distinctions between ground and excited states.

**Authors:**

**Dahbia Talbi (Univ. of Montpellier – CNRS, Fr)**

**Jean-Christophe Loison (Univ. of Bordeaux – CNRS, Fr)**

**$C(^3P) + NH_2(^2B_1)$ → $HCN(X\ ^1\Sigma^+) + H(^2S)$ (1)** $\Delta Hr_{298} = -554$ kJ/mol refs (2) and (3)

**→ $HNC(X\ ^1\Sigma^+) + H(^2S)$ (2)** $\Delta Hr_{298} = -499$ kJ/mol refs (2) and (3)

## Rate Coefficient Data ($k = k_1 + k_2$)

| $k$ / cm$^3$ molecule$^{-1}$ s$^{-1}$ | $T$ / K | Reference | Comments |
|---|---|---|---|
| *Rate Coefficient Measurements* | | | |
| No data was found | | | |
| *Calculations* | | | |
| $k = 2.3\ 10^{-10}$ | 10 | Herbst et al. (2000) | |
| $k = 1.6\ 10^{-10}$ | 20 | Herbst et al. (2000) | |
| $k = 6.8\ 10^{-11}$ | 300 | Herbst et al. (2000) | |
| *Reviews and Evaluations* | | | |
| $k_1$ (T) = 3.26 x $10^{-11}$ $(T/300)^{-0.36}$ | 10-300 | UMIST database | |
| $k_2$ (T) = 3.26 x $10^{-11}$ $(T/300)^{-0.36}$ | 10-300 | UMIST database | |
| $k_1$ (T) = 3.26 x $10^{-11}$ $(T/300)^{-0.36}$ | | OSU database | |
| $k_2$ (T) = 3.26 x $10^{-11}$ $(T/300)^{-0.36}$ | | OSU database | |

## Comments

To our knowledge no experimental studies have been undertaken to determine rate coefficients for these reactions. The only known reaction rates are from the dynamical calculations of Herbst et al (1). Using the potential energy surface calculated by D. Talbi & Y Ellinger (2) by means of accurate ab initio methods, Herbst et al. (1) have determined the rate coefficients given above. The dynamical study has also revealed that the products (HCN and HNC) are formed with so much excess of energy that efficient isomerization occurs leading to an equal production rate for HNC and HCN from both reactions.

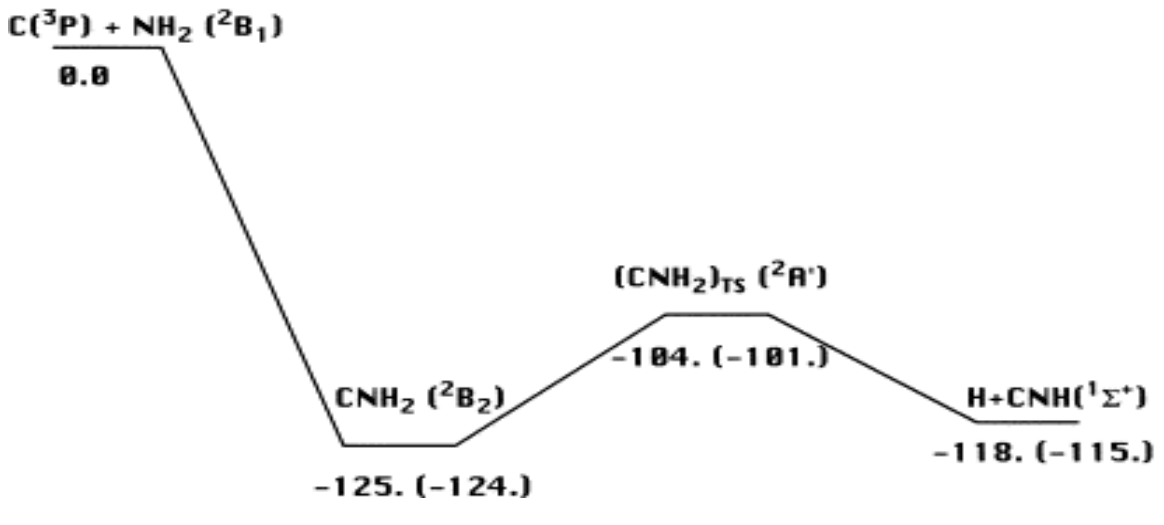

The Energy profiles are for surfaces of doublet multiplicity. Relative energies, given in kcal/mol, have been calculated at the PMP4SDTQ/6-311++G(3df,3pd)//MP2/6-31G(d,p) and CCSD(T)/6-311++G(3df,3pd)//MP2/6-31G(d,p) (numbers in brackets) levels. In all cases, relative energies are corrected for the ZPE and for spin contamination from higher spin state

## Preferred Values

*Rate coefficient (10 – 300 K)*

$\mathbf{k_1}$ **(T) = $3\times10^{-11}(T/300)^{-0.2}e^{-6/T}$ cm$^3$s$^{-1}$**

$\mathbf{k_2}$ **(T) = $3\times10^{-11}(T/300)^{-0.2}e^{-6/T}$ cm$^3$s$^{-1}$**

*Reliability*

**$F_0$ = 1.5 ; g = 0**

*Comments on Preferred Values*

**Author:**

**Ian Smith (University of Cambridge, UK)**

**$O(^3P) + C_3N(X^2\Sigma^+) \rightarrow CO(^1\Sigma^+) + C_2N(^2\Pi)$ (1)** No thermochemical data for $C_3N$

## Rate Coefficient Data *k*

| $k$ / cm$^3$ molecule$^{-1}$ s$^{-1}$ | $T$ / K | Reference |
|---|---|---|
| *Rate Coefficient Measurements* | | |
| No measurements in the literature were found. | | |
| *Reviews and Evaluations* | | |
| $4 \times 10^{-11}$ | 10 – 300 | UMIST database |
| $4 \times 10^{-11}$ | no $T$-dependence | OSU website |

## Comments

This radical-radical reaction (forming CO) is presumably exothermic and is spin-allowed (over doublet PESs). However, the reactants also correlate with quartet PESs.

## Preferred Values

*Rate coefficient (10 – 300 K)*

$k$(300 K) = $1 \times 10^{-10}$ cm$^3$ molecule$^{-1}$ s$^{-1}$
$k$(10 K) = $1 \times 10^{-10}$ cm$^3$ molecule$^{-1}$ s$^{-1}$
**$k$(T) = $1 \times 10^{-10}$ cm$^3$ molecule$^{-1}$ s$^{-1}$**

*Reliability*

$\Delta \log k$ (300 K) = ± 0.5
$\Delta \log k$ (10 K) = ± 0.6
**$F_0 = 3$ ; $g = 2.97$**

*Comments on Preferred Values*

The UMIST and Ohio databases adopt the same rate coefficient value – though it is not clear how they arrive at this value. It seems to me to be slightly low – even allowing for the non-reactive quintet PESs that correlate with the reactants.

## References

**Authors:**

**Ian Smith (University of Cambridge, UK)**

**Jean-Christophe Loison (Univ. of Bordeaux - CNRS, Fr)**

**Stephen Klippenstein (Argonne National Laboratory, Argonne, IL, USA)**

$$O(^3P) + CN(X^2\Sigma^+) \rightarrow N(^4S) + CO(X^1\Sigma^+) \quad (1)\ \Delta Hr^{298} = -\ 322.4\ kJ\ mol^{-1}$$ *(Baulch* et al.*, 2005)*

$$\rightarrow N(^2D) + CO(X^1\Sigma^+) \quad (2)\ \Delta Hr^{298} = -\ 92.3\ kJ\ mol^{-1}$$ *(Baulch* et al.*, 2005)*

## Rate Coefficient Data ($k = k_1 + k_2$)

| $k$ / $cm^3$ $molecule^{-1}$ $s^{-1}$ | $T$ / K | Reference | |
|---|---|---|---|
| *Rate Coefficient Measurements* | | | |
| $k = (10.5 \pm 5.8) \times 10^{-11} \cdot \exp(-1200 \pm 350)/T)$ | 570-687 | Boden and Thrush, 1968 | (2) |
| $k = 2.0 \times 10^{-11}$ | 298 | Schacke et al. 1974 | (3) |
| $k = 2.0 \times 10^{-11}$ | 298-387 | Albers et al. 1975 | (4) |
| $k = 1.7 \pm 0.7 \times 10^{-11}$ | 298 | Schmatjko and Wolfrum, 1977 | (5) |
| $k = 1.8 \times 10^{-11}$ | 295 | Schmatjko and Wolfrum, 1978 | (6) |
| $k = 3.1\ (+2.6/-1.3) \times 10^{-11}$ | 2000 | Louge and Hanson, 1984 | (7) |
| $k = 13.0 \cdot \pm 2.6$ | 3000-4500 | Davidson et al., 1991 | (8) |
| $k = 3.69 \pm 0.75 \times 10^{-11}$ | 298 | Titarchuk and Halpern, 1995 | (9) |
| *Theory* | | | |
| $k = 4.35 \times 10^{-11} \cdot (T/298)^{0.46} \cdot \exp(-364/T)$ | 300-5000 | Cobos, 1996 (Statistical) | (10) |
| $k = 1.42 \times 10^{-10} \cdot (T/298)^{0.13} \cdot \exp(-5.3/T)$ | 5-400 | Andersson, 2003 (QCT) | (11) |
| $k = 8.69 \times 10^{-10} \cdot (T/300)^{0.17} / (5+3\exp(-288/T)+\exp(-326/T)$ | 15-400 | Klippenstein, 2011 (TST) | |

## Comments

There have been several scattered measurements of the rate coefficients for this reaction. We believed that the determinations at room temperature (3-6,9) are the more accurate. Studies (3-6) all used a combined discharge flow/flash photolysis method (O-atoms from discharge through $He/O_2$ mixture, CN radicals were produced from photolysis of $C_2N_2$) and O atomss (in excess /CN) by microwave discharge. These measurements seem reliable but the authors give few details on how they measure/estimate the [O] atom concentration. Moreover, they have to deal with CN vibrational relaxation and secondary N + CN reaction, with N issued from O + CN. In the paper by Albers et al. (4), they report a $T$-independent rate coefficient for O + CN. But their value of the rate coefficient for CN + $O_2$ (which should be much easier to determine than those for O + CN) reaction is a factor of about 2.5 smaller than the currently accepted – and very well determined – value, casting some doubt on their measurements on O + CN. Study (9) used a double photolysis technique (CN from BrCN and $O(^3P)$ from $SO_2$ or $N_2O$, CN being detected directly by LIF). In this last study the O-atom concentrations were estimated from the estimated photon flux and the absorption cross-sections. This method, using absolute cross section absorption of precursor, is not straightforward. However, we attach more weight to this last determination and we estimate an "average" value of the experimental rate coefficient at room temperature equal to $k(298K) = 3 \times 10^{-11}$ $cm^3$ $molecule^{-1}$ $s^{-1}$. No measurements of low-temperature rate coefficients have been reported.

It should be noted that both channels are exothermic and allowed by spin-orbit correlation rules. The fact that more surfaces correlate with the reactants than pass adiabatically to the products, suggests that the rate coefficient should be lower than a simple collisional estimate. Reaction (2) to $CO(^1\Sigma^+)$ + $N(^2D)$ on doublet surfaces is likely to proceed through strongly bound states of NCO ($X^2\mathbf{\Pi}$, $A^2\Sigma^+$, $B^2\mathbf{\Pi}$) and is likely to be favoured over reaction (1) via quartet surfaces to $CO(^1\Sigma^+)$ + $N(^4S)$. (On these surfaces, there is no initial 'pairing' of electrons from each of the radical reactants.) This conclusion is supported by theoretical work by Abrahamsson *et al.* (12) which suggests that there is a substantial barrier on the lowest quartet surface – at least for linear configurations. Previous ab-initio calculations from the same team (11) using two potential energy surfaces corresponding to electronic states of $^2A'$ and $^2A''$ symmetry lead to potential barriers on both surfaces for the nonlinear approach of O toward CN. This makes the rate of reaction decrease with decreasing temperature below 200 K. Moreover, it is almost exclusively N + CO that is formed due to the C + NO channel being endothermic by about 1.2 eV. This means that trajectories entering (near-) collinearly into the CNO minimum emerge as nonreactive in most cases. Quasi-classical trajectory calculations lead to a rate that is constant not in good agreement with most experimental rate coefficients (fig 10 of Andersson et al. (11)):

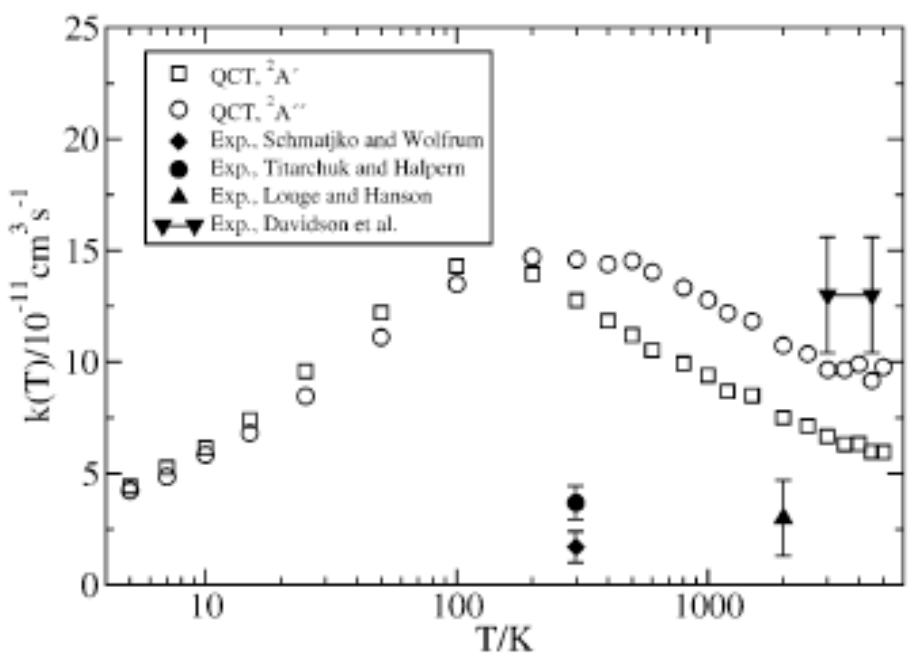

**Figure 10.** Total thermal rate coefficients for the O + CN reaction. Open symbols correspond to calculated rate coefficients for the $^2A'$ and $^2A''$ surfaces, and filled symbols correspond to experimental results (refs 35−38).

As pointed by the authors, is it likely that the theoretical rate coefficients for the O + CN presented in the paper are too high in the low-temperature regime. So the T dependence of this reaction is difficult to estimate. There is likely a competition between two opposite effects: increase at low temperature due to the rather strong, attractive, long-range forces between the dipole moment of CN and the quadrupole moment of O($^3$P) atoms and the presence of a barrier for the nonlinear approach of O toward CN, which decreases the rate coefficient at low temperature (the electronic degeneracy factor due to $O_J$ population varying only from 0.30 at 300 K to 0.40 at 10K). The statistical works of Cobos (10) seems not to be reliable as he found a barrier leading to an unexpected very low rate coefficient at low temperature. In 2011 Stephen Klippenstein performed (for this datasheet) TST calculations leading to k(CN+O) = $1.68\times10^{-10}T^{1/6}$ [2/(5+3exp(-288/T)+exp(-326/T)], in relatively good agreement with Anderson *et al.* calculations (11), and so notably higher than experimental determinations. No clear explanation has been found to explain this difference.

For the branching ratio between (1) and (2), Wolfrum et al have determined a N($^2$D)/N($^4$S) branching ratio of 5.7 using a bi-modal CO(v) distribution. They interpret the bi-modal distribution in terms of different energy distribution pathways for the two N($^2$D) and N($^4$S) channels, with the distribution for the N($^4$S) channel being inverted and non-statistical. This measurement is questionable as first the quadruplet surfaces are supposed to be repulsive, and also because their directly measurement of a N($^2$D)/N($^4$S) branching ratio of 25 using VUV absorption (using discharge lamp) is very different, and no convincing explanation is provided (they suggest that N($^4$S) reacts with $C_2N_2$ but but in general N($^2$D) has a higher reactivity than N($^4$S)). However, as the radiative lifetime of the $^2$D excited state of N atoms is *ca.* 26 hours ($A$ ~ $7.4\times10^{-6}$ s$^{-1}$), spontaneous radiation will be the main loss process for N($^2$D) where the gas density is less than *ca.* $10^6$ cm$^{-3}$ (as in most regions of the ISM).

## Preferred Values

*Total rate coefficient (10 – 300 K)*

$\mathbf{k_1 + k_2 = 5\times10^{-11}}$ **cm$^3$ molecule$^{-1}$ s$^{-1}$**

*Branching Ratios*

$\mathbf{k_1 / (k_1 + k_2) = 0.15}$

$\mathbf{k_2 / (k_1 + k_2) = 0.85}$

*Reliability*

$\mathbf{F_0 = 3,\ g = 0}$

*Comments on Preferred Values*

It's difficult to recommend an overall rate coefficient as the experimental rate coefficients are scattered at 298 K and the two calculations are 3 to 10 times higher than the experimental values. As we don't see any evidence to choose a particular value, we propose an intermediate value between experimental and theoretical ones of $5.0\times10^{-11}$ cm$^3$ molecule$^{-1}$ s$^{-1}$ with an uncertainty of a factor of 3 for the entire range between 10 and 300 K.

**Author:**

**Ian Smith (University of Cambridge, UK)**

$$O(^3P) + HNO(^1A') \rightarrow NO_2(^2A_1) + H(^2S) \quad \textbf{(1)}\ \Delta Hr^{298} = -105.3\ \text{kJ mol}^{-1} \quad \textit{(Baulch et al., 2005)}$$

$$\rightarrow NO(^2\Pi) + OH(^2\Pi) \quad \textbf{(2)}\ \Delta Hr^{298} = -227.2\ \text{kJ mol}^{-1} \quad \textit{(Baulch et al., 2005)}$$

$$\rightarrow O_2(^3\Sigma_g^-) + NH(^3\Sigma^-) \quad \textbf{(3)}\ \Delta Hr^{298} = +0.54\ \text{kJ mol}^{-1} \quad \textit{(Baulch et al., 2005)}$$

## Rate Coefficient Data $k = k_1 + k_2 + k_3$

| $k$ / cm$^3$ molecule$^{-1}$ s$^{-1}$ | $T$ / K | Reference | Comments |
|---|---|---|---|
| *Rate Coefficient Measurements* | | | |
| $k = 3.8 \times 10^{-11}$ | 242-473 | (a) Inomata and Washida. | O-atoms produced in flow-discharge, $NH_2$ from laser photolysis of $NH_3$. $NH_2$ and HNO concentrations followed by photoionisation mass spec. |
| *Reviews and Evaluations* | | | |
| $k_1 = 1 \times 10^{-12}$ | 10 – 300 | UMIST database | |
| $k_2 = 6 \times 10^{-11}$ | 10 – 2500 | UMIST database | |
| $k_3$ negligible | 116 – 300 | UMIST database | |
| $k_1 = 1 \times 10^{-12}$ | no $T$-dependence | OSU website | |
| $k_2 = 3.8 \times 10^{-11}$ | no $T$-dependence | OSU website | |

## Comments

The reactions via channels (1) and (2) are strongly exothermic and spin-allowed. It should be noted that HNO is a singlet in its electronic ground state, so the main reaction, (2), is an H-atom abstraction (rather than one proceeding through an adduct). The experimental measurement in (a) seems reliable and provides an overall rate coefficient. There is some theoretical evidence (b) to support the idea that (2) is the dominant channel. Inomata and Washida (a) find no $T$-dependence over the range studied. The UMIST and OSU data bases assume a very small contribution from channel (1) but the grounds for this are unclear.

## Preferred Values

### *Rate coefficient (10 – 300 K)*

$k$(300 K) = 3.8x10$^{-11}$ cm$^3$ molecule$^{-1}$ s$^{-1}$
$k$(10 K) = 5x10$^{-11}$ cm$^3$ molecule$^{-1}$ s$^{-1}$
**$k(T) = 3.8\times10^{-11}\ (T/300)^{-0.08}$ cm$^3$ molecule$^{-1}$ s$^{-1}$**

### *Branching Ratios*

$\mathbf{k_2 / (k_1 + k_2 + k_3) = 1.0}$
$\mathbf{k_1 / (k_1 + k_2 + k_3) = k_3 / (k_1 + k_2 + k_3) = 0.0}$

### *Reliability*

$\Delta \log k$ (300 K) = ± 0.3
$\Delta \log k$ (10 K) = ± 0.6
$\mathbf{F_0 = 2\ ;\ g = 7}$

### *Comments on Preferred Values*

For $k$(300 K), the value of $k$ measured in (a) is recommended. The $T$-dependence is hard to predict. I have assumed no barrier and allowed for a small negative $T$-dependence at lower temperatures.

**Author:**

**Ian Smith (University of Cambridge, UK)**

$$\mathbf{O(^3P) + NH(^3\Sigma^-) \rightarrow NO(^2\Pi) + H(^2S)} \quad \mathbf{(1)} \quad \Delta Hr^{298} = -\ 297.4\ \text{kJ mol}^{-1} \quad (*)$$

$$\mathbf{\rightarrow OH(^2\Pi) + N(^4S)} \quad \mathbf{(2)} \quad \Delta Hr^{298} = -\ 95.9\ \text{kJ mol}^{-1} \quad (*)$$

## Rate Coefficient Data $k = k_1 + k_2$

| $k$ / cm$^3$ molecule$^{-1}$ s$^{-1}$ | $T$ / K | Reference | Comments |
|---|---|---|---|
| *Rate Coefficient Measurements* | | | |
| $6.6 \times 10^{-11}$ | 295 | Adamson *et al.*, 1994 | (a) |
| $k_2 < 1.66 \times 10^{-13}$ | 298 | Hack *et al.*, 1994 | (b) |
| *Reviews and Evaluations* | | | |
| $1.8 \times 10^{-10} \exp(-300/T)$ | 295 – 3500 | Baulch *et al.*, 2005 | (*) |
| $k(298\ K) = 6.7 \times 10^{-11}$; $k_2\ (298\ K) < 1.7 \times 10^{-13}$ | | | |
| $k_1 = k_2 = 1.16 \times 10^{-10}$ | 250 – 3000 | UMIST database | |
| $k_1 = k_2 = 1.16 \times 10^{-10}$ | all temperatures | OSU website | |

## Comments

Channel (1) is strongly exothermic and could occur via the ground ($^1A'$) state of HNO (and possibly excited states). The reactants correlate with 27 states ($^5A'$ + $2^5A''$, $^3A'$ + $2^3A''$, $^1A'$ + $2^1A''$), the products with 8 states ($^3A'$ + $^3A''$, $^1A'$ + $^1A''$). Therefore, there is an electronic degeneracy factor of *ca*. 8/27.

There are scarcely any kinetic experiments on this reaction. The principal aim in the experiments described in (a) was to find the rate coefficient for O + $NH_2$. However, the interpretation of the observations to yield the rate coefficient for O + NH is quite direct and appears sound. Hidden in the text the authors propose a branching ration into channel (2) of 7%. They also refer to an earlier measurement by Wagner's group in fair agreement with their value. Ref. (b) reports a very low branching ratio to channel (2) – in agreement with its lower exothermicity and the notion that reaction may occur *via* HNO.

## Preferred Values

### *Rate coefficients (10 – 300 K)*

$k = k_1$ (298 K) = 6.6 · $10^{-11}$ cm$^3$ molecule$^{-1}$ s$^{-1}$

$k = k_1$ (10 K) = 6.6 · $10^{-11}$ cm$^3$ molecule$^{-1}$ s$^{-1}$

**$k_1(T)$ = 6.6 · $10^{-11}$ cm$^3$ molecule$^{-1}$ s$^{-1}$**

**$k_2$ (298 K) = $k_2$ (10 K) = zero**

### *Reliability*

$\Delta \log k$ (300 K) = ± 0.5

$\Delta \log k$ (10 K) = ± 0.6

**$F_0$ = 3 ; g = 2.97**

### *Comments on Preferred Values*

The value recommended for $k = k_1$ (298 K) is about what one would get by reducing a collisional rate coefficient by the factor of 8/27. I have assumed no temperature-dependence. I also believe that the branching ratio to channel (2) is likely to be small in agreement with the measurement in (b). I don't know where the values in the data bases come from. I recommend values that are lower by a factor of *ca*. 2.

**Author:**

**Ian Smith (University of Cambridge, UK)**

$$O(^3P) + NH_2(^2B_1) \rightarrow HNO(^1A') + H(^2S) \quad (1) \quad \Delta Hr^{298} = -113.7 \text{ kJ mol}^{-1} \quad (*)$$

$$\rightarrow NO(^2\Pi) + H_2(^1\Sigma_g^+) \quad (2) \quad \Delta Hr^{298} = -348.4 \text{ kJ mol}^{-1} \quad (*)$$

$$\rightarrow OH(^2\Pi) + NH(^3\Sigma^-) \quad (3) \quad \Delta Hr^{298} = -45 \text{ kJ mol}^{-1} \quad (*)$$

## Rate Coefficient Data $k = k_1 + k_2 + k_3$

| $k$ / cm$^3$ molecule$^{-1}$ s$^{-1}$ | $T$ / K | Reference | Comments |
|---|---|---|---|
| *Rate Coefficient Measurements* | | | |
| $k = 7.6 \times 10^{-11}$ | 296 | (a) Dransfield *et al.* | all studies use variants of the flow-discharge method and should be quite reliable – in respect of the overall rate coefficient |
| $k = 6.5 \times 10^{-11}$ | 295 | (b) Adamson *et al.* | |
| $k = (1.2 \pm 0.3) \times 10^{-11}$ | 242 - 473 | (c) Inomata and Washida | |
| *Reviews and Evaluations* | | | |
| $4.56 \times 10^{-11} \exp(10/T)$ | 200 – 3000 | UMIST database | |
| $8 \times 10^{-11}$ | no $T$-dependence | OSU website | |

## Comments

This radical-radical reaction is strongly exothermic and spin-allowed to all channels – though the reactants correlate with quartet, as well as doublet, surfaces.
The values of the overall rate coefficient at 'room temperature' obtained in (a) and (b) are in good agreement; that in (c) is slightly higher but probably within combined errors. Consequently, $k$ can be assumed to be quite well-determined. Reaction probably proceeds through an $ONH_2^*$ complex whose formation does not require passage over a barrier.
Refs (a) and (b) also agree that (1) is the major channel with $k_3$ being about 10% of $k$. There is no experimental indication that channel (2) proceeds at a measurable rate.

## Preferred Values

### *Total rate coefficient (10 – 300 K)*

$k(300\text{ K}) = 7.0 \cdot 10^{-11}$ cm$^3$ molecule$^{-1}$ s$^{-1}$
$k(10\text{ K}) = 1.0 \cdot 10^{-10}$ cm$^3$ molecule$^{-1}$ s$^{-1}$
**$k(T) = 7 \times 10^{-11} (T/300)^{-0.1}$ cm$^3$ molecule$^{-1}$ s$^{-1}$**

### *Branching Ratios*

**$k_1 / (k_1 + k_2 + k_3) = 0.9$**
**$k_2 / (k_1 + k_2 + k_3) = 0.0$**
**$k_3 / (k_1 + k_2 + k_3) = 0.1$**

### *Reliability*

$\Delta \log k$ (300 K) = ± 0.3
$\Delta \log k$ (10 K) = ± 0.5
**$F_0 = 2$ ; g = 4**

### *Comments on Preferred Values*

I recommend values for the overall rate coefficient similar to those in the UMIST and Ohio compilations. For $k$(298 K), I take an average of the room temperature measurements in (a) and (b). I assume a mild negative $T$-dependence.
If all three channels proceed via formation of an energised $ONH_2$ complex, then the branching ratios are unlikely to have a strong temperature-dependence.

**Author:**

**Ian Smith (University of Cambridge, UK)**

| | | | |
|---|---|---|---|
| **$CH(^2\Pi) + NH_3$ → $H_2CNH + H$** | **(1)** | $\Delta H_r^{298} = -\ 243.8$ kJ mol$^{-1}$ |
| **→ $CH_2 + NH_2$** | **(2)** | $\Delta H_r^{298} = +\ 29.4$ kJ mol$^{-1}$ |
| **→ $CH_3 + NH$** | **(3)** | $\Delta H_r^{298} = -\ 47.6$ kJ mol$^{-1}$ |
| **→ $HCN + H_2 + H$** | **(4)** | $\Delta H_r^{298} = -\ 197.4$ kJ mol$^{-1}$ |
| **→ $HCNH_2 + H$** | **(5)** | $\Delta H_r^{298} = -\ 89.5$ kJ mol$^{-1}$ |

The enthalpies of reaction are those given by Blitz et al. (4). See also Baulch et al. (*).

## Rate Coefficient Data ($k = k_1 + k_2 + k_3 + k_4 + k_5$)

| $k$ / cm$^3$ molecule$^{-1}$ s$^{-1}$ | $T$ / K | Reference | |
|---|---|---|---|
| *Rate Coefficient Measurements (k)* | | | |
| $(8.6 \pm 0.6) \times 10^{-11} \exp\{(230 \pm 30)/T\}$ | 297 − 677 | Zabarnick, Fleming, Lin, 1989 | (1) |
| $(7.2 \pm 1.7) \times 10^{-11} \exp\{(317 \pm 13)/T\}$ | 300 − 1300 | Becker, et al 1993 | (2) |
| $1.69\mathrm{x}10^{-10}$ (T/298 K)$^{-0.56}$ exp(-28 /T) | 23 − 295 | Bocherel *et al.*, 1996 | (3) |
| $(1.5 \pm 0.05)\ \mathrm{x}10^{-10}$ | 298 | Blitz et al. 2012 | (4a) |
| $(1.25 \pm 0.2)\ \mathrm{x}10^{-10}$ | 298 | Blitz et al. 2012 | (4b) |

*Branching Ratios*

From their combined experimental and theoretical study, Blitz et al (4) inferred that ca. 96% of the reaction proceeds to $H_2CNH$ + H (that is, by channel (1)), Approximately 4% may proceed via channel (3).

| | | | |
|---|---|---|---|
| *Reviews and Evaluations* | | | |
| $k_1 = 1.69 \times 10^{-10}$ (T/300 K)$^{-0.41}$ exp(−19.0 / T) | 23 − 1300 | UMIST/UDFA database | |
| no values given | | OSU website | |

## Comments

Refs. (1), (2) and (3): all these studies used the reliable pulsed photolysis / laser-induced fluorescence (PP/LIF) method. They yield very similar values for $k_1$ at 298 K. By extension the low $T$ measurements reported in (3) can be considered reliable. The expression for the temperature-dependence of the overall rate coefficient given in the UMIST/UDFA data base appears to be based on these three studies and is similar to that given in ref. (3).

In ref. (4), Blitz et al. measured the rate coefficient at 298 K both by the PP/LIF method (4a) and by observing the rise in the LIF signals from the H atom product (4b). Both results are in good agreement with the earlier values at 298 K.

However, the main purpose of the study of Blitz et al. was to determine the branching ratios for this reaction. Their calculations (ab initio, transition state theory, and master equation) indicated that channel (1) is dominant. The experiments showed that the H atom yield from the reaction is 0.89 ± 0.07

## Preferred Values

### *Rate coefficients (10 – 300 K)*

$k$(300 K) = 1.6 · $10^{-10}$ cm$^3$ molecule$^{-1}$ s$^{-1}$

$k$(10 K) = $1.9\mathrm{x}10^{-10}$ cm$^3$ molecule$^{-1}$ s$^{-1}$

**$k(T) = 1.6\mathrm{x}10^{-10}\ (T/300)^{-0.05}$ cm$^3$ molecule$^{-1}$ s$^{-1}$**

### *Branching ratios*

**$k_1 / (k_1 + k_2 + k_3 + k_4 + k_5) = 0.95 \pm 0.05$**

**$k_3 / (k_1 + k_2 + k_3 + k_4) = 0.05 \pm 0.05$**

### *Reliability*

**$F_0 = 1.2$ g = 2**

### *Comments on Preferred Values*

Given the good agreement between the experimental values at 298 K, the estimate of 20% certainty seems generous. The wider uncertainty at 10 K reflects the fact that the measurements in (3) only go down to 25 K and it is not clear if $k_1$ will continue to increase below 25 K.

**Author:**

**Ian Smith (University of Cambridge, UK)**

**$CN + NH_3 \rightarrow HCN + NH_2$** **(1)** $\Delta Hr^{298} = -65.4$ kJ mol$^{-1}$ (*)

**$\rightarrow NCNH_2 + H$** **(2)** $\Delta Hr^{298} = -49$ kJ mol$^{-1}$ (*)

## Rate Coefficient Data ($k = k_1 + k_2$)

| $k$ / cm$^3$ molecule$^{-1}$ s$^{-1}$ | $T$ / K | Reference | Comments |
|---|---|---|---|
| *Rate Coefficient Measurements (k)* | | | |
| $8.8 \times 10^{-12}$ | 687 | Boden and Thrush, 1968 | |
| $(2.5 \pm 0.5) \times 10^{-11}$ | 295 | DeJuan, Smith and Veyret | |
| $(1.52 \pm 0.23) \times 10^{-11} \exp\{(1.50 \pm 0.40)$ kJ/mole/RT$\}$ | 294 - 761 | Sims and Smith | |
| $(2.9 \pm 0.3) \times 10^{-11}$ | 296 | Meads, Maclagan and Phillips, 1993 | |
| $(2.8 \pm 0.7) \times 10^{-11}$ (T/298 K)$^{(-1.14\pm0.15)}$ | 25 - 295 | Sims *et al.*, 1994 | |
| *Branching Ratios* | | | |
| $k_1 / (k_1 + k_2) = 1.0$; $k_2 / (k_1 + k_2) = 0.0$ | | Talbi and Smith, 2009 | |
| $k_1 / (k_1 + k_2) = 1.0$; $k_2 / (k_1 + k_2) = 0.0$ | | Blitz, Seakins and Smith, 2009 | |
| *Reviews and Evaluations* | | | |
| $k_1 = 3.41 \times 10^{-11}$ (T/300 K)$^{-0.90}$ exp(9.9 K / T) | all temperatures | UMIST database | |
| $k_2 = 1.38 \times 10^{-11}$ (T/300 K)$^{-1.14}$ | all temperatures | UMIST database | |
| $k_1 = 1.38 \times 10^{-11}$ (T/300 K)$^{-1.14}$ | all temperatures | OSU website | |
| $k_2 = 1.3 \times 10^{-11}$ (T/300 K)$^{-1.14}$ | all temperatures | OSU website | |

## Comments

Refs (b), (c) and (e): All these studies used the reliable pulsed photolysis / laser-induced fluorescence method. They yield very similar values for $k_1 + k_2$ at 298 K, which agree with that from Meads *et al.* So the value of $k_1 + k_2$ at 298 K must be judged well-established. By extension the low $T$ measurements reported in (c) can be considered reliable.

Until recently, there was really no evidence as to the major products of this reaction (i.e., the branching ratio between channels (1) and (2)). The UMIST and OSU databases appear to arrive at individual values of $k_1$ and $k_2$ by arbitrarily dividing up the overall rate coefficient between the two channels. Meads *et al.* (d) demonstrated the formation of $NH_2$ but their other efforts to find products did not eliminate channel (2). Nor, unfortunately, did their *ab initio* calculations cast a clear light on this problem.

Recently there have been theoretical (f) and experimental (g) studies of this reaction with the emphasis on determining the branching ratio between reactions (1) and (2). Talbi and Smith (f) found no low energy path to $NCNH_2$ + H and concluded that reaction proceeds exclusively to HCN + $NH_2$. Likewise, the experiments of Blitz *et al.* (g) found no significant formation to H-atoms and they concluded that reaction must proceed only via channel (1).

## Preferred Values

### *Rate coefficients (10 – 300 K)*

$k$(300 K) = 2.8x10$^{-11}$ cm$^3$ molecule$^{-1}$ s$^{-1}$
$k$(10 K) = 5x10$^{-10}$ cm$^3$ molecule$^{-1}$ s$^{-1}$
**$k_1(T)$ = 2.8x10$^{-11}$ ($T$/300)$^{-0.85}$ cm$^3$ molecule$^{-1}$ s$^{-1}$**

### *Branching ratios*

**$k_1 / (k_1 + k_2) = 1.0$**
**$k_2 / (k_1 + k_2) = 0.0$**

### *Reliability*

$\Delta \log k$(298 K) = ± 0.08
$\Delta \log k$(10 K) = ± 0.15

**$F_0 = 1.2$ ; $g = 1.6$**

### *Comments on Preferred Values*

Given the good agreement between the experimental values at 298 K, the estimate of 20% certainty seems generous. The wider uncertainty at 10 K reflects the fact that the measurements in (e) only go down to 25 K and it is not clear if $\{k_1 + k_2\}$ will continue to increase below 25 K. The experiments of Blitz et al. (h) show that $k_2$ /

($k_1$ + $k_2$) is certainly less than 0.05, and most probably zero.

**Author:**

**Ian Smith (University of Cambridge, UK)**

$$\mathbf{H_3^+ + N(^4S) \rightarrow NH_2^+ + H} \quad \mathbf{(1)} \quad \Delta Hr^{298} = -\ 98 \text{ kJ mol}^{-1} \qquad \text{refs 1 and 3}$$

$$\mathbf{\rightarrow NH^+ + H_2} \quad \mathbf{(2)} \quad \Delta Hr^{298} = +\ 82 \text{ kJ mol}^{-1} \qquad \text{refs 1 and 3}$$

**Rate Coefficient Data *k***

| $k$ / $cm^3$ $molecule^{-1}$ $s^{-1}$ | $T$ / K | Reference | Comments |
|---|---|---|---|
| *Rate Coefficient Measurements* | | | |
| $(4.5 \pm 1.8)x10^{-10}$ | 295 ± 5 | Scott et al. (1997) | |
| $< 5x10^{-11}$ | 295 ± 5 | Milligan et al. (2000) | |
| too small to estimate | all temperatures | Bettens & Collins (1998) | |
| *Reviews and Evaluations* | | | |
| this reaction is not included | | UMIST database | |
| $1x10^{-17}$ (i.e., effectively zero) | | OSU website | |

## Comments

Reference (1): Experiments using a selected ion flow tube. Because $N_2$ reacts rapidly with $H_3^+$ and the fraction of N atoms produced from $N_2$ by microwave discharge is small, a relative rate measurement is adopted by measuring the yields of $NH_4^+$ and $N_2H^+$.
Reference (2): Measurements made by the same group as in (1) with an improved flowing afterglow/selected ion flow tube apparatus. No evidence was found for the formation of $NH_2^+$ or $NH_3^+$ and the upper limit for the rate coefficient was derived. A probable source of error in the earlier experiments was given.
Reference (3): A detailed theoretical study using an *ab initio* potential energy surface and quasiclassical trajectories. No trajectories leading to $NH_2^+ + H_2$ were found. The possibility of a non-adiabatic mechanism was discussed and rejected.

## Preferred Values

This reaction is too slow to influence chemistry in cold ISCs and can be omitted from models.

### *Reliability*

### *Comments on Preferred Values*

Channel (2) is strongly endothermic. Theory (both (3) and earlier work by (4)) were unable to find a low energy path for channel (1). It seems safe to assume that this reaction will not occur in interstellar clouds. Other routes for the formation of $NH_3$ must exist.

**Authors:**

**Jean-Christophe Loison (Univ. of Bordeaux - CNRS, Fr)**

**Dahbia Talbi (Univ. of Montpellier – CNRS, Fr)**

# $NH_3^+(X^2A''_2) + H_2(X^1\Sigma_g^+) \rightarrow NH_4^+(X\ ^1A_1) + H(^2S)$ $\Delta Hr^{298} = -87$ kJ mol$^{-1}$ (*)

## Rate Coefficient Data $k$

| $k$ / cm$^3$ molecule$^{-1}$ s$^{-1}$ | $T$ / K | Reference | Comments |
|---|---|---|---|
| *Rate Coefficient Measurements (k)* | | | |
| $5 \times 10^{-13}$ | 300 | Kim *et al*, 1975 | (2) |
| $1.7 \times 10^{-11} \times \exp(-1200/T)$ | 300-800 | Fehsenfeld *et al*, 1975 | (3) |
| $(2.09 \pm 0.22) \times 10^{-11}$ | 85-510 | Adams and Smith, 1984 | (4) |
| Complex dependency | 20-298 | Borhinger, 1985 | (5) |
| $3.36\times10^{-14}\times\exp(35.7/T)$ | 11-20 | Barlow and Dunn, 1987 | (6) |
| *Theory* | | | |
| Complex dependency | 5-350 | Herbst *et al*, 1991 | (1) |
| *Reviews and Evaluation* | | | |
| $3.36\text{x}10^{-14}\exp(35.7/T)$ | 10-20 | UMIST database | |
| $2.0\text{x}10^{-13}$ | 20-300 | UMIST database | |
| $1.7\text{x}10^{-11}\exp(-1044/T)$ | 300-41000 | UMIST database | |
| $1.5\text{x}10^{-14}(T/300)^{-1.5}$ | | OSU database | |

## Comments

There have been various experimental investigations of this reaction. Original studies by Kim *et al.* and Fehsenfeld *et al* showed the reaction to be quite slow at room temperature ($k = 5 \times 10^{-13}$ cm$^3$ molecule$^{-1}$ s$^{-1}$) and to increase in rate with increasing temperature, indicating a system with an activation energy of approximately 2.1 kcal mol$^{-1}$ (8.8 kJ mol$^{-1}$). Subsequent experimental work by Adams and Smith indicated however that the rate coefficient does not obey a typical Arrhenius relation at temperatures significantly under 300 K but rather levels off at temperatures near 80-100 K. Finally, work by Luine and Dunn and Bohringer showed that the rate coefficient actually increases at still lower temperatures.

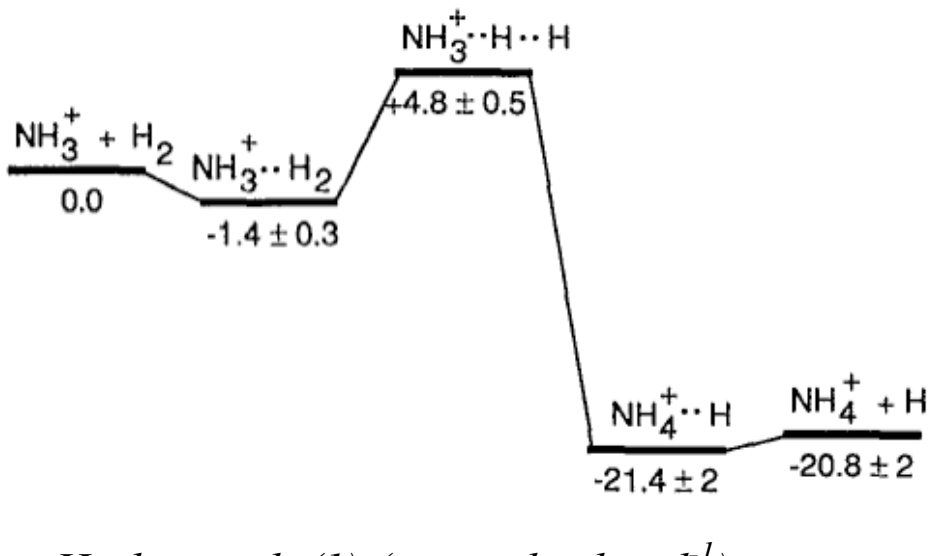

*from Herbst et al. (1) (unit in kcal mol$^{-1}$)*

At low temperatures, the lifetime of the complex is quite long and even a small tunneling probability can be important. Barlow and Dunn extended the work of Luine and Dunn and studied the effects of deuterium substitution on the rate of the reaction at low temperature. Their results strongly support the tunneling hypothesis.

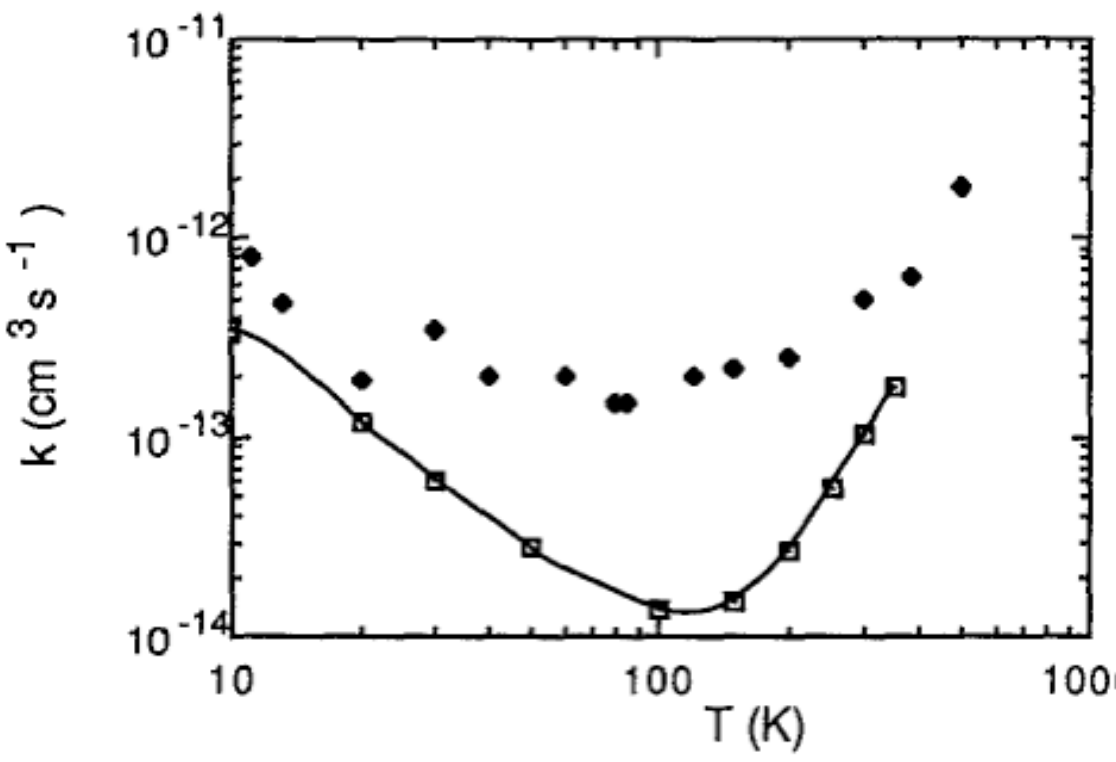

*Experimental rate coefficients (filled diamonds) and calculated one (open square). from Herbst et al. (1)*

The rate coefficients calculated using a phase space theory approach (1) where the formation of an initial complex is considered, and from which tunneling under a small transition state barrier could occur are lower than the measured values, even when large uncertainties are considered for the calculated barrier height and imaginary vibrational frequency. The lower calculated value for the rate coefficient may be due either to the fact that in the experimental determination, part of the reaction is due to three-body mechanism or to the fact that the rate of tunneling have been calculated using an Eckart potential, which may lead to substantial error as the tunneling is notoriously sensitive to details of the potential surface such as the height and imaginary vibrational frequency which regulate the steepness of the barrier.

## Preferred Values

*Rate coefficient (10 – 20 K)*
**$k$ (T) = $3.36\times10^{-14}\times\exp(35.7/T)$ cm$^3$ molecule$^{-1}$ s$^{-1}$**
*Rate coefficient (20 – 230 K)*
**$k$ (T) = $2.0\times10^{-13}$ cm$^3$ molecule$^{-1}$ s$^{-1}$**
*Rate coefficient (230 – 800 K)*
**$k$ (T) = $1.7\times10^{-11}\times\exp(-1044/T)$ cm$^3$ molecule$^{-1}$ s$^{-1}$**

*Reliability*
**$F_0$ = 1.5, g = 0**

*Comments on Preferred Values*
There are some uncertainties at low temperature if the three-body mechanism plays an important role.

**Authors:**

**Dahbia Talbi (Univ. of Montpellier – CNRS, Fr)**

**Wolf Geppert (Stockholm University, Se)**

| | | | |
|---|---|---|---|
| $\mathbf{N_2H^+ + e^- \rightarrow N_2 + H}$ | **(1)** | $\Delta Hr^{298} = -\ 815.3$ kJ mol$^{-1}$ | ref (3) |
| $\mathbf{\rightarrow NH + N}$ | **(2)** | $\Delta Hr^{298} = -\ 117.1$ kJ mol$^{-1}$ | ref (3) |

## Rate Coefficient Data $k = k_1 + k_2$

| $k$ / cm$^3$ molecule$^{-1}$ s$^{-1}$ | $T$ / K | Reference |
|---|---|---|
| *Rate Coefficient Measurements* | | |
| $k(T) = 7.5 \times 10^{-7}$ | 300 | (1) |
| $k(T) = 1.7 \times 10^{-7} (T/300)^{-0.9}$ | 95 – 300 | (2) |
| $k(T) = (1.0 \pm 0.1) \times 10^{-7} (T/300)^{-0.5 \pm 0.02}$ | 10 – 300 | (3) |
| $k(T) = 2.4 \times 10^{-7}$ | 300 | (4) |
| $k(T) = 2.98 \times 10^{-7} (T/300)^{-0.74}$ | 10 – 150 | (6) |
| $k(T) = 2.74 \times 10^{-7} (T/300)^{-0.84}$ | 150 – 1000 | (6) |
| *Branching Ratios* | | |
| $k_1 / (k_1 + k_2) = 0.36$ ; $k_2 / (k_1 + k_2) = 0.64$ | | (3) |
| $k_1 / (k_1 + k_2) \approx 0.95 \pm 0.02$; $k_2 / (k_1 + k_2) \approx 0.05 \pm 0.02$ | | (5) |
| $k_2 / (k_1 + k_2) = 0.08$ | | (6) |
| *Reviews and Evaluations* | | |

## Comments

The first measurement of recombination rate coefficient for $N_2H^+$ was done by Mul and MacGowan (1) using a mearged-beam technique. They determined at 300 K a rate coefficient of 7.5 x$10^{-7}$ $cm^3s^{-1}$. Smith and Adams(2) using a flowing afterglow technique found a lower rate of 1.7 x $10^{-7}$ $cm^3s^{-1}$. Later on, Geppert *et al.* (3) at CRYRING obtained a close value of 1 x $10^{-7}$ $cm^3s^{-1}$ for the same temperature. More recently Poteyra et al. (4) with a revisited flowing afterglow technique measured for 300 K a slightly higher rate of 2.4 x $10^{-7}$ $cm^3s^{-1}$. In the new experiment by Vigren et al. a similar reaction rate of 2.98 x $10^{-7}$ $cm^3s^{-1}$ was determined fror T = 300K.

Geppert et al. (3) also determined a branching ratio with the CRYRING experiment and found that the NH + H channel was much larger than previously believed i.e 64% for NH + N and 36% for $N_2$ + H, but a later experiment (6) showed that this unexpected branching ratio was due to a contamination of the ion beam and a new branching ration was proposed (6) with for the NH + H channel a value of 7 %. This is in good agreement with the branching ratio measured by Molek *et al.* (5) of less than 5% for the NH + N channel and the theoretical investigations of D. Talbi (7,8,9) showing from potential energy surface calculations for linear $N_2H$ and $N_2H^+$ that the likely outcomes from the dissociative recombination $N_2H^+$ are the $N_2$ and H fragments, with $N_2$ in its first electronically excited state, while the NH + N channel should be minor because of inefficient curves crossing.

Since the rate constants of the dissociative recombination of $N_2H^+$ measured by the last flowing afterglow and storage ring experiments do not differ very much we recommend an intermediate value of $k(T) = (2.6 \pm 0.6) \times 10^{-7}(T/300)^{-0.84}$. For the branching ratio we also choose a value in agreement with both studies, namely 5 ± 2 % for the NH + N and 95 ± 2 % for the $N_2$ + H channel.

## Preferred Values

*Total rate coefficient (10 – 1000 K)*

$$k(T) = 2.6 \times 10^{-7}(T/300)^{-0.84}$$

*Branching ratios*

$k_1 / (k_1 + k_2) = 0.95$

$k_2 / (k_1 + k_2) = 0.05$

## Reliability

$F_0 = 1.6$, $g = 0$

## Comments on Preferred Values